\documentclass[12pt,twocolumn,a4paper]{article}

\usepackage{times}
\usepackage[numbers]{natbib}
\usepackage{graphicx}

\usepackage{authblk}

\title{World ships: Feasibility and Rationale}
\date{}

\author[1]{Andreas Makoto Hein\thanks{Corresponding author. E-mail: andreas.hein@i4is.org}}
\author[2]{Cameron Smith}
\author[3]{Fr\'ed\'eric Marin}
\author[4]{Kai Staats}

\affil[1]{\small Initiative for Interstellar Studies, 27-29 South Lambeth Road, London SW8 1SZ, United Kingdom}
\affil[2]{\small Department of Anthropology, Portland State University Portland, OR, 97207, USA}
\affil[3]{\small Universit\'e de Strasbourg, CNRS, Observatoire Astronomique de Strasbourg, UMR 7550, F-67000 Strasbourg, France}
\affil[4]{\small Arizona State University Interplanetary Initiative, Phoenix, Arizona, USA}

\begin{document}

\maketitle

\begin{abstract}
World ships are hypothetical, large, self-contained spacecraft for crewed interstellar travel, 
taking centuries to reach other stars. Due to their crewed nature, size, and long trip times, 
the feasibility of world ships faces an additional set of challenges compared to interstellar 
probes. Despite their emergence in the 1980s, most of these topics remain unexplored. This article 
revisits some of the key feasibility issues of world ships. First, definitions of world ships 
from the literature are revisited and the notion of world ship positioned with respect to similar 
concepts such as generation ships. Second, the key question of population size is revisited 
in light of recent results from the literature. Third, socio-technical and economic feasibility issues are evaluated. Finally, world ships are compared to potential alternative modes of crewed interstellar travel. Key roadblocks for world ships are the considerable resources required, shifting its economic feasibility beyond the year 2300, and the development of a maintenance system capable of detecting, replacing, and repairing several components per second. The emergence of alternative, less costly modes of crewed interstellar travel at an earlier point in time might render world ships obsolete. 
\end{abstract}

\section{Introduction} \label{S1}

\begin{table*}[t]
  \caption{Crewed starship categories with respect to cruise velocity and population size.}
  \centering
  \begin{tabular}{p{0.2\linewidth}p{0.2\linewidth}p{0.2\linewidth}p{0.2\linewidth}}
  \hline
  & & \textbf{Population size} & \\
  \textbf{Cruise velocity} [\%c] & \textbf{$<$ 1000} & \textbf{$<$ 100,000} & \textbf{$>$ 100,000} \\
  \hline
  $>$ 10 & Sprinter & Colony ship & -\\
  $<$ 10 & Slow boat & Colony ship & World ship\\
  $<$ 1 & - & Colony ship & World ship\\
  \hline
  \end{tabular}
  \label{Tab1}  
\end{table*}

\begin{figure*}[t]
  \centering
  \includegraphics[width=0.9\linewidth]{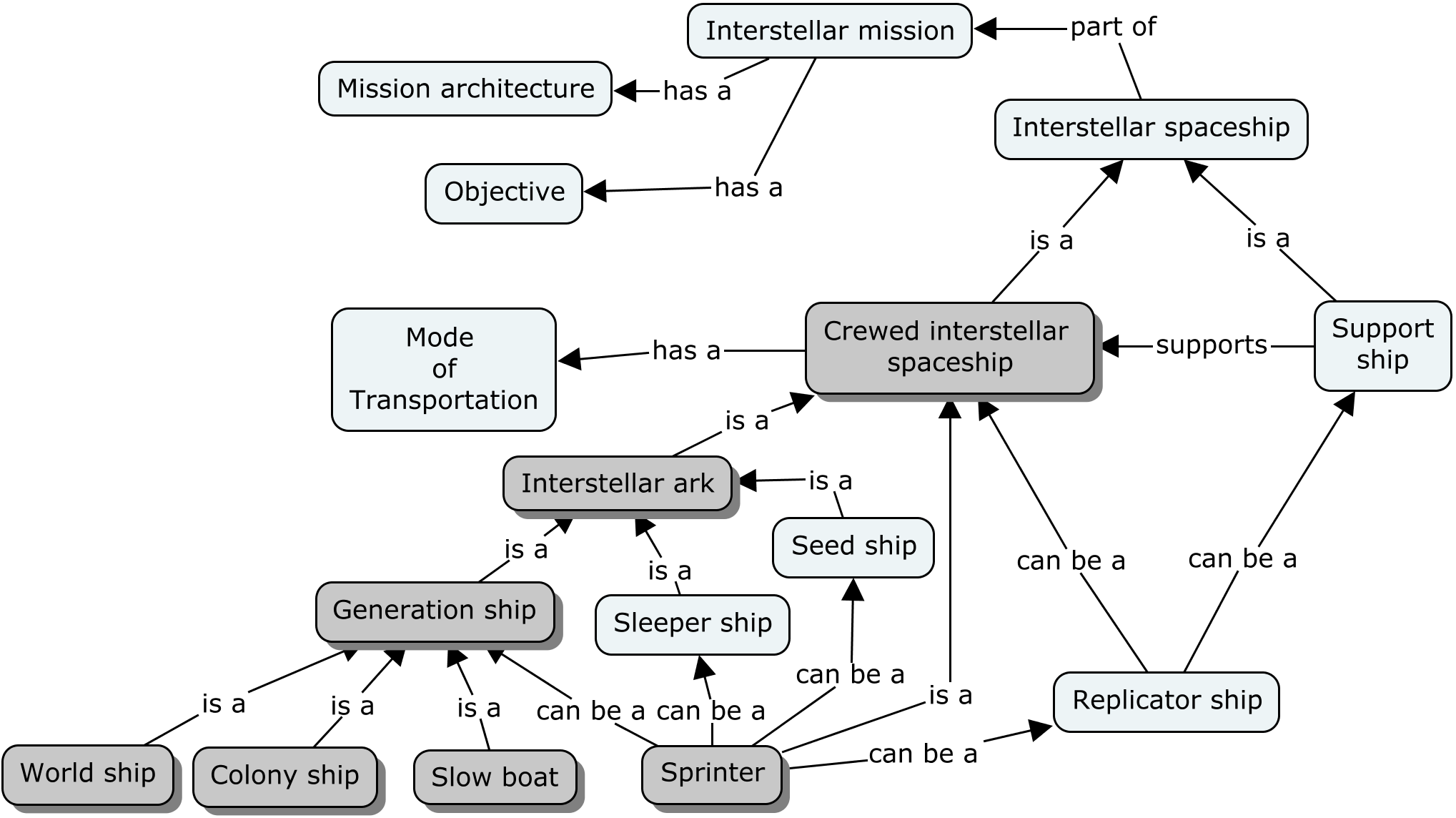}
  \caption{Concept map for crewed interstellar starships.}
  \label{Fig1}
\end{figure*}

World ships are hypothetical large, self-contained, self-sufficient crewed spacecraft for interstellar travel. Large, artificial habitats appeared in the literature as early as 1929 in Bernal's "The World, the Flesh and the Devil" \cite{Bernal2018}.
However, the notion was extensively discussed for the first time in a special issue of the Journal of the British Interplanetary Society (BIS) in 1984. 
Martin \cite{Martin1984} characterizes a world ship as a “large, lumbering vehicle, moving at a fraction of a 
per cent of the speed of light and taking millennia to complete a journey between stars.” Martin \cite{Martin1984} 
presents a rationale for world ships, cost estimates, and how scenarios for their construction and colonization might 
look like. In the special issue, Bond and Martin \cite{Bond} present an analysis of engineering feasibility, 
including two types of world ships, each with a different (land or sea-based) environment for its crew to live 
in. Grant \cite{Grant1984} goes on to analyze the stability of isolated world ship populations and fleets, and 
Smith \cite{Smitha} provides scenarios for how life on a world ship might look like. Finally, Holmes \cite{Holmes} provides a sociological perspective on world ships and how isolated communities could be sustained over millennia. The special issue's articles conclude that there is no fundamental technical, economic, or social reason which
would prohibit the construction of world ships. However, due to their mass on the order of billions of tons, their 
construction is estimated to take place several centuries in the future \cite{Martin1984}, when humanity would have 
control over solar system resources.\\
\indent Apart from world ships, Finney and Jones \cite{Finney1986} and Kondo \cite{Kondo2003} have explored in their edited
volumes the idea of generation ships in general, with contributions covering technical, cultural, and social aspects. \\
\indent In 2011, a World Ship Symposium was organized by the BIS, resulting in another world ship JBIS special issue in 2012, 
including contributions dealing with the shift from a planetary to a space-based civilization \cite{Ashworth2012,Ashworth2012a}, 
financing such projects \cite{Ceyssens2012}, and propulsion systems \cite{Matloff2012a}. Notably, Hein et al. Ceyssens et al. \cite{Ceyssens2012} analyzed how a world ship project might be funded and proposes a long-term investment 
approach in which funds are accumulated over centuries. Hein et al. \cite{Hein2012b} 
provide a reassessment of world ship feasibility, taking additional aspects such as knowledge transfer and reliability into 
account. Furthermore, a fundamental trade-off between trip duration and population size is hypothesized, as longer trip durations require a larger population number for sustaining the required skillset. From a reliability perspective, it is concluded that an extremely complex technical system such as a world ship would require a sophisticated maintenance system, as the number of components that would need to be replaced, repaired, or both amounts to several per second. \\
\indent Some of the team members who worked on this paper subsequently founded Project Hyperion in the context of Icarus Interstellar. Within Project Hyperion, Smith \cite{Smith2014b} published a seminal article on the required population size for a world ship for trip times of several centuries. He concludes that a population size that takes genetic drift and 
catastrophic events into account would comprise several tens of thousands of people. The paper received a lot of attention,
in particular, as it contradicts previous population estimates, which were much lower, such as just a few crew members in 
Finney and Jones \cite{Finney1986}. It also confirms that longer trip times correlate with larger population sizes. 
More recently, a team lead by Marin presented a further analysis 
of population size, in which much smaller population sizes are again obtained \cite{Marin2017,Marin2018,Marin2018a}.\\
\indent Apart from population size estimations, world ships have been treated in dedicated workshop tracks at the Tennessee Valley Interstellar Workshop (TVIW) in 2016 and 2017, emphasizing ecological engineering issues of world ships \cite{Cobbs2015}. Furthermore, in 2015, a student team at the International Space University (ISU) has developed the "Astra Planeta" concept for a world ship, covering a wide range of topics, such as technical, legal, societal aspects, as well as governance and financing \cite{Acierno2015}. \\
\indent This paper provides an updated overview of research on world ships. It covers some crucial topics such as how to define world ships, population size, socio-technical and economic feasibility, and how world ships might fit into the broader landscape of crewed interstellar travel concepts.  

\section{Revisiting definitions}

An attempt to distinguish between different concepts for crewed interstellar travel was provided in Hein et al. \cite{Hein2012b}. 
The distinction is made based on two dimensions: cruise velocity and population size. Crewed starships with populations 
below 1000 and a velocity higher than 10\% of the speed of light are called ``sprinter'', slower starships with a similar crew 
size ``slow boat'' and starships with a population size below 100,000 are called ``colony ship''. World ships are defined as crewed starships with populations over 100,000 and a velocity below 10\% of the speed of light. As a result, we get the following three criteria, adapted from \cite{Hein2012b}: 

\begin{itemize}
\item Self-sufficiency: thousands of years
\item Population size: $>$ 100,000
\item Cruise velocity: $<$ 1\%c
\end{itemize}

An overview of these concepts
is shown in Table ~\ref{Tab1} and Fig. ~\ref{Fig1}. Fig. ~\ref{Fig1}, in particular, shows a concept map for crewed interstellar spacecraft 
from Hein et al. \cite{Hein2012b}. It can be seen that all four concepts of crewed starships (sprinter, slow boat, colony ship, 
world ship) are generation ships and also considered interstellar arks. \\
\indent There are several assumptions behind this taxonomy and concept map. First, the population size should be taken as order of magnitude
values and are somewhat arbitrary. One could draw an alternative demarcation line at one million between colony ships and world 
ships. Hence, it might be better to rather speak of a fuzzy demarcation line between these concepts. For example, in Hein et al. 
\cite{Hein2012b}, a world ship design is presented, based on several stacked Stanford Tori. The Stanford Torus was imagined for 
population sizes of about 10,000 to 100,000, but it can be seen that by stacking a sufficient number of Tori, a population of 100,000 to one million can 
be accommodated without fundamentally changing the nature of the spacecraft. Second, the velocity range of world ships is larger 
than in Martin \cite{Martin1984}, extending velocities to below 0.1$c$, as there is no physical or engineering reason why world 
ship velocities should be limited to below 0.01$c$. As mentioned in the original paper, the most crucial parameter is trip time, which we would consider at least on the order of centuries. \\
\indent A more fundamental issue with the existing definitions is that they do not explicitly reflect on the meaning of "world" in "world ship." A "world" goes beyond self-sufficiency and a given population size. "World" commonly denominates Earth with all life and human civilization. If this is what we mean by "world" in "world ship," any spacecraft with a closed habitat containing life and a human civilization could be called "world ship." However, this interpretation of "world" has the connotation of a habitat with an enormous size. We can even imagine a habitat the size of a planet, along with the living conditions on a planet. We will later present such a planet-sized world ship, based on the McKendree Cylinder in Section 3. The etymology of "world" allows for an alternative interpretation, where "world" indicates a material universe or ontology. A "world ship" would then be a ship which, for humans on-board, would represent "all there is," not only in a material sense (what is inside the habitat, spacecraft subsystems, etc.) but also in terms of what humans would conceive as the "reality" in which they live. Hence, departing from the existing definitions in the literature, interesting new interpretations of world ships are possible, going back to the meaning of "world."    
\section{World ship designs}\label{S3}
\begin{figure*}[t]
  \centering
\includegraphics[width=1\linewidth]{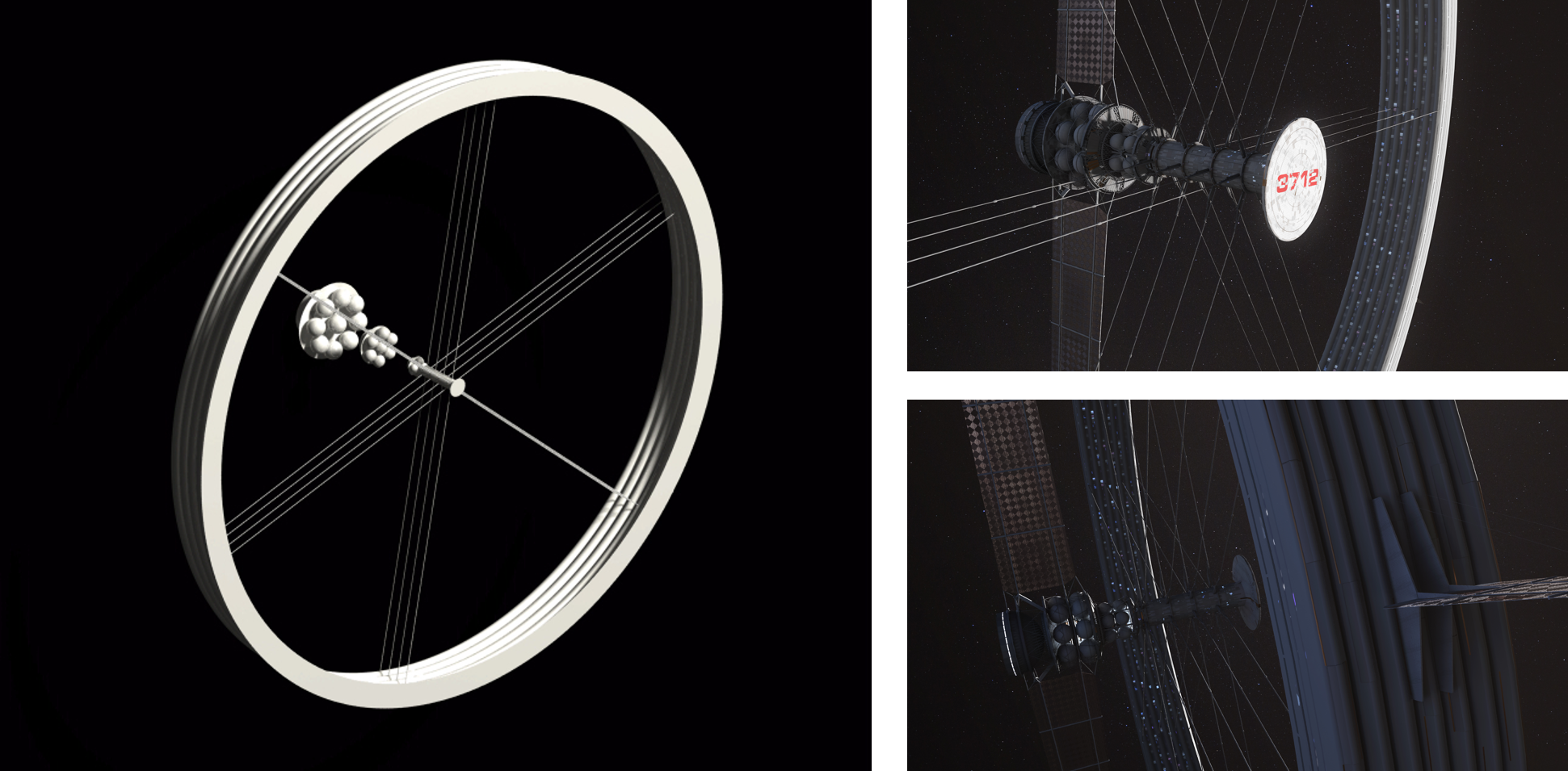}
    \caption{World ship based on the Deadalus fusion propulsion system and stacked Stanford Tori \cite{Hein2012b}. Artistic impressions by Adrian Mann (left) and Maciej Rebisz (right).}
  \label{Fig3}
\end{figure*}
\begin{figure*}[t]
  \centering
  \includegraphics[width=0.9\linewidth]{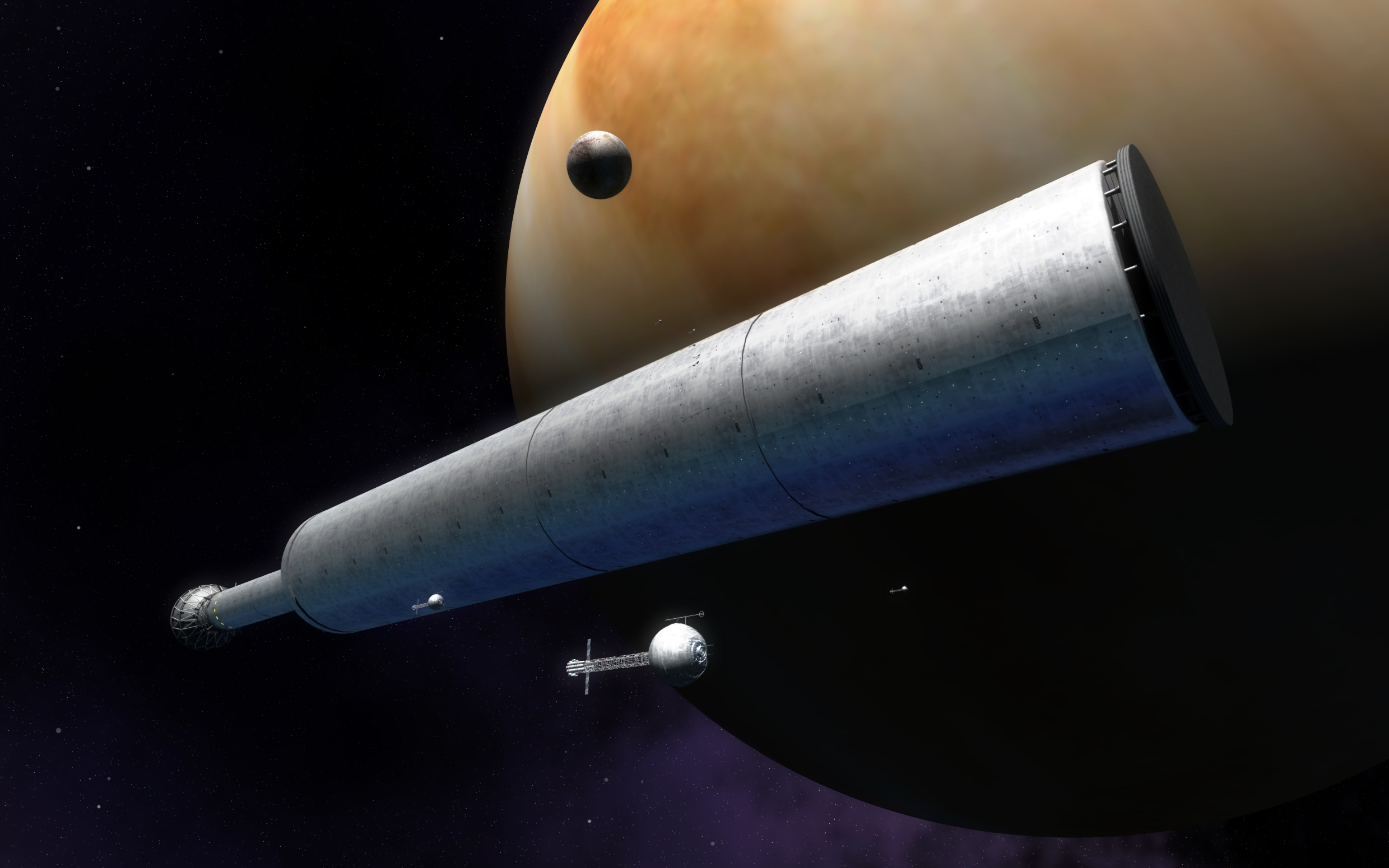}
  \caption{``Wet World'' type world ship (image credit: Adrian Mann).}
  \label{Fig4}
\end{figure*}
\begin{figure*}[t]
  \centering
  \includegraphics[width=0.9\linewidth]{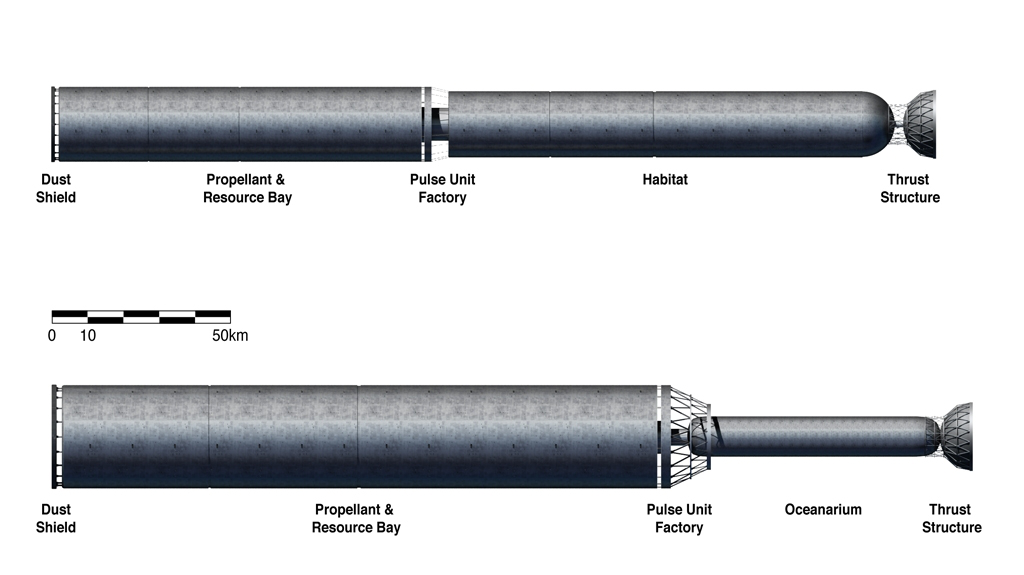}
  \caption{Size comparison of ``Dry World'' and ``Wet world'' 
	   type world ships (image credit: Adrian Mann).}
  \label{Fig5}
\end{figure*}
World ships designs are usually dominated by a large habitat section and a comparatively small propulsion section. All other 
subsystems of crewed spacecraft are also present, however, their size is much smaller, compared to the habitat and propulsion subsystem. Only few engineering designs of world ships have been presented in the literature. Matloff \cite{Matloff1976} presented 
a world ship based on an O'Neill ``Model 1'' colony \cite{ONeill1974} in 1976. Two cylindrical habitats are attached to the propulsion
system, which is placed between them. A Deadalus-type fusion propulsion system is used. Power is provided by fusion reactors. 
Deceleration is taking place via an electric sail. O'Neill himself proposed the use of an O'Neill colony with an antimatter propulsion
system as a world ship \cite{ONeill1977}. Such world ships would gradually move out of the Solar System and embark on an interstellar
trip. However, except for the propellant mass, no details about the design were given. \\
\indent More recently Hein et al. \cite{Hein2012b} have presented a world ship design with stacked Stanford Tori for population sizes 
on the order of $10^4$ to $10^5$, shown in Fig.~\ref{Fig3}. Similar to \cite{Matloff1976}, this world ship design is based 
on the Daedalus fusion propulsion system and a habitat design borrowed from an O'Neill colony, in this case the Stanford Torus. 
Fig.~\ref{Fig3} also shows the dust shield put on top of the Stanford Torus facing flight direction. The authors of \cite{Hein2012b} have 
subsequently further developed the design, in order to reduce the overall mass of the spacecraft, which is dominated by the shielding 
mass for the habitat ($>$90\%). One possibility would be to use the deceleration propellant as the shielding material. The propellant
mass mainly consists of Deuterium, which has similar shielding characteristics as hydrogen \cite{Simnad2001}. The propellant is used up during the last
years of the trip for decelerating the spacecraft and would serve as a shielding up to this point. The two disadvantages of this 
approach are that the complexity of the spacecraft increases. The fuel needs to be transported from the shield to the fusion engine. 
The fuel pellets either need to be manufactured on-board or the shielding is already in the form of fuel pellets. In both cases 
additional equipment has to be installed. \\
\indent The Bond and Martin \cite{Bond} world ships are the largest proposed world ships in terms size and mass. Figs.~\ref{Fig4} and \ref{Fig5} show a artist conceptions of the Bond and Martin world ships \cite{Bond}. In particular 
Fig.~\ref{Fig5} provides a size comparison of the "Dry World" (habitable area is mostly land) and "Wet World" (habitable area 
is mostly water) world ships. It can be seen that the habitat (large cylindrical section) and the propulsion module (second 
cylinder in the back with nozzle) are dominating the designs. In addition, a flat, circular dust shield is attached to the front 
of the world ship. \\
\indent Although the Bond and Martin world ships are the largest proposed world ships in the literature, even larger world ships can be imagined, such as world ships based on the McKendree Cylinder, with a length of 4610 km, a radius of 461 km, a mass of $8.0*10^{16} kg$, and a population of $99*10^{12}$, about 12 times the current human population on Earth \cite{McKendree1996}. A world ship of that size would be in principle feasible, given the resources in the Solar System \cite{McKendree1996}. \\
\indent Table \ref{wsvalues} provides an overview of key parameter values of world ship designs in the literature.
The population size and cruise velocity is of the same order of magnitude for all designs. However, there are orders of magnitude differences for the dry mass and propellant mass. These differences are a result of different assumptions regarding the size of the habitat. The Bond and Martin world ships are replicating living conditions in sparsely settled areas on Earth. The Stanford Torus rather replicates an urban or suburban area with a comparatively high population density. Finally, the Enzman starship seems to rather replicate a high-density urban area. The population size is assumed to increase 10 times during the trip. For the habitat mass, equipment,  and consumables, a mass between 150 t/person at the beginning of the journey and 15 t/person at the end is assumed.
\begin{table*}[t]
\caption{World ship designs from the literature with key values}
\label{wsvalues}
\centering
\begin{tabular}{p{0.35\linewidth}p{0.12\linewidth}p{0.12\linewidth}p{0.12\linewidth}p{0.12\linewidth}}
\hline
\textbf{Design} & \textbf{Popula- tion size} & \textbf{Dry mass [tons]} & \textbf{Propellant mass [tons]} & \textbf{Cruise velocity [\%c]} \\
\hline
\textbf{Enzman world ship \cite{Crowl2012a}} & 20,000 - 200,000 & 300,000 & $3 \cdot 10^6$ &  0.9\\
\textbf{Torus world ship \cite{Hein2012b}} & 100,000 & $10^7$ & $5 \cdot 10^7$ &  1 \\
\textbf{Dry world ship - Mark 2A \cite{Bond}} & 250,000 & $2.0\cdot10^{11}$ & $8.2\cdot10^{11}$ &  0.5\\
\textbf{Dry world ship - Mark 2B \cite{Bond}} & 250,000 & $5.7\cdot10^{11}$ & $2.3\cdot10^{12}$ &  0.5\\
\textbf{Wet world ship \cite{Bond}} & 250,000 & $2.2\cdot10^{12}$ & $9.0\cdot10^{12}$ &  0.5\\
\hline
\end{tabular}
\end{table*}
To conclude, existing world ship designs are based on a fusion propulsion system and a large habitat. The habitat size and mass depends on the underlying assumptions about the environment in which the crew would live. 

\section{World ship feasibility criteria} \label{S4}
In the following, we decompose world ship feasibility into biological, cultural, social, technical, and economic criteria. Biological feasibility includes the genetic health of the population during the trip and at the point where they start a new settlement at the target destination. Hence, an important precondition for a world ship, we must assume that habitats in which human populations can live out multiple generations can be constructed. These will be informed by decades of life in other beyond-Earth settlements, such as Mars and / or orbital communities, such as described in \cite{Hein2012b}. Studies in closed-system ecology are underway or have been demonstrated to some extent with Biosphere-2 or BIOS-3. We understand genetic health as states of being adapted to a set of environmental factors well enough to ensure successful self-replication. Cultural feasibility includes how knowledge is transferred and preserved, including knowledge which is essential for living on the world ship and starting a settlement at the target destination. Social feasibility includes, but is not limited to criteria that are related to the organization of the society on-board of a world ship, such as its stability. Technical criteria are related to the technologies used on a world ship, their maturity and performance. Economic criteria are related to the economic preconditions that allow for the development of a world ship, such as the scope of economic activities and wealth. Apart from analyzing these feasibility criteria in isolation, we will also look into areas where feasibility criteria depend on each others. Notably, we look into socio-technical feasibility. In addition, we will compare world ships to alternative ways of crewed interstellar travel. This point seems important to us, as world ships will not get built if faster, cheaper, and less risky ways of interstellar travel are going to be developed. Table \ref{Taba} provides an overview of these feasibility criteria. 
\begin{table*}[t]
  \centering
  \caption{Overview of world ship feasibility criteria and their impact on key design parameters}
  \begin{tabular}{p{0.3\linewidth}p{0.3\linewidth}p{0.3\linewidth}}
  \hline
  \textbf{Feasibility category} & \textbf{Criteria} & \textbf{Design considerations}\\
  \hline
    \textbf{Biological} & Genetics & population size, trip duration  \\ 
    \textbf{Cultural} & Knowledge transmission & population size, knowledge management approach \\ 
    \textbf{Social} & Societal structure & habitat geometry, size, modularity \\ 
    \textbf{Technical} & Performance of technologies & Velocity, trip duration \\ 
    ~ & Maturity of technologies & Precursors  \\ 
    ~ & Reliability of technologies & Spare parts mass, maintenance system \\ 
    \textbf{Economic} & Scope of economic activities & Scope of materials \\ 
    ~ & Wealth & Affordable size, mass \\ 
  \hline  
  \end{tabular}
  \label{Taba}  
\end{table*}

\section{Biological and cultural feasibility - World ship population estimations}\label{S5}

\subsection{World ship population and composition: time and space boundaries}
The project of interstellar voyaging is ultimately meant to preserve and spread human life in space, an idea which is  rooted in various cultural traditions, ranging from the 'Great Navigator' in Polynesian culture to 'leaving thd cradle' narrative by Konstantin Tsiolkovsky. Therefore it builds out from 
the central concerns of the human body. For the exploratory period of short-term spaceflight, the concerns of the individual 
body were measured in days and months, or up to a year. These are the scales of biology and flight physiology. As we move towards 
consideration of permanent space settlement and even interstellar voyaging to exoplanets, concern must expand to include issues of
individual bodies arranged as families, families arranged as communities, communities as a population (a ‘deme'), and populations 
as cultures. These are the special domain of demography, population genetics and the scientific study of humanity, anthropology.
In particular, anthropology studies human biocultural evolution as humanity adapts both by gene and (moreso in the last 100,000 years) 
culture. \\
\indent Determination of a world ship population depends on the objective. Our objective is to allow a genetically- and culturally-healthy 
population to arrive at an exoplanet, where they may land and begin a new world for humanity. 14 stars
are within 10 light years of Earth; propulsion engineering and other issues related to the feasibility of reaching each are explored elsewhere 
\cite{Long2016}. Alternative destinations are introduced later in Section \ref{S61}. Exoplanet discoveries are burgeoning, with a measured number of 3971 exoplanets discovered 
as of January the 23rd, 2019 (see http://http://exoplanet.eu/). The current paradigm is that there are ``2 $\pm$ 1 planets in the habitable
zone of each [Milky Way galaxy] star'' \cite{Boivard}. The closest habitable planet may well be within 10 light years of Earth. We will
likely know within a few decades as new exoplanet characterization tools are developed; these are scheduled to include the James Webb Space 
Telescope \cite{Beichman2014}, the Extremely Large Telescope and the ExoLife Finder \cite{Berdyugina2018}. \\
\indent The 10 light years distance is selected as a boundary here for reasons of time and space. It represents a distance just reasonably 
possible to reach, with reasonably-expectable world ship engineering, in several centuries if reasonably-expectable propulsion speeds 
are achieved \cite{Heller2017,Diverse2012}. If the objective is to land a healthy population of humans (and their many domesticates) 
on an exoplanet after some centuries, we must know how many humans are required to establish a new population that itself will be 
multigenerationally viable. A number of such estimates have been made since the 1980's and are discussed below.

\subsection{Biological health: estimates of world ship populations to date}
\begin{table*}[t]
  \centering
  \caption{Seven published and/or current world ship population estimates. Results of SIMOC, 
	   noted in the final row, are imminent and not yet published. D1 is the recommended 
	   interstellar emigrant founding population. D2, mentioned in the text, is not 
	   noted in this table.}
  \begin{tabular}{p{0.13\linewidth}p{0.15\linewidth}p{0.16\linewidth}p{0.20\linewidth}p{0.25\linewidth}}
  \hline
  \textbf{Model} & \textbf{Model type} & \textbf{Spacefaring simulations?} & \textbf{D1} & \textbf{Current regard} \\
  \hline
  ETHNOPOP & Demographic & Few & $<$300 & likely low \\
  \hline
  SMITH & Statistical & Several & $>$7,500, ideally larger (14,000 -- 44,000) & lower \& middle  figures reasonable, higher figures too high \\
  \hline
  GARDNER-O'KEARNY & Statistical Agent-Based & Several & $>$2,000 & possibly reasonable \\
  \hline
  HERITAGE (paper 1) & Monte Carlo Agent-Based & Many, ongoing & $>$5,000 & possibly reasonable \\
  \hline
  HERITAGE (paper 2) & Monte Carlo Agent-Based & Many, ongoing & mathematical minimum 98, ideally larger & possibly reasonable \\
  \hline  
  HERITAGE (paper 3) & Monte Carlo Agent-Based & Many, ongoing & circa 500 & possibly reasonable \\
  \hline  
  HERITAGE + SMITH & Monte Carlo Agent-Based + Anthropological & Yes, ongoing & Some multiples of 500 -- 1,000 person “village modules” & biologically and culturally realistic and reasonable \\
  \hline  
  SIMOC & Monte Carlo Agent-Based & Many (parallel computing) & Unknown & Unknown \\
  \hline
  \end{tabular}
  \label{Tab2}  
\end{table*}
\begin{figure*}[!t]
  \centering
  \includegraphics[width=0.8\linewidth]{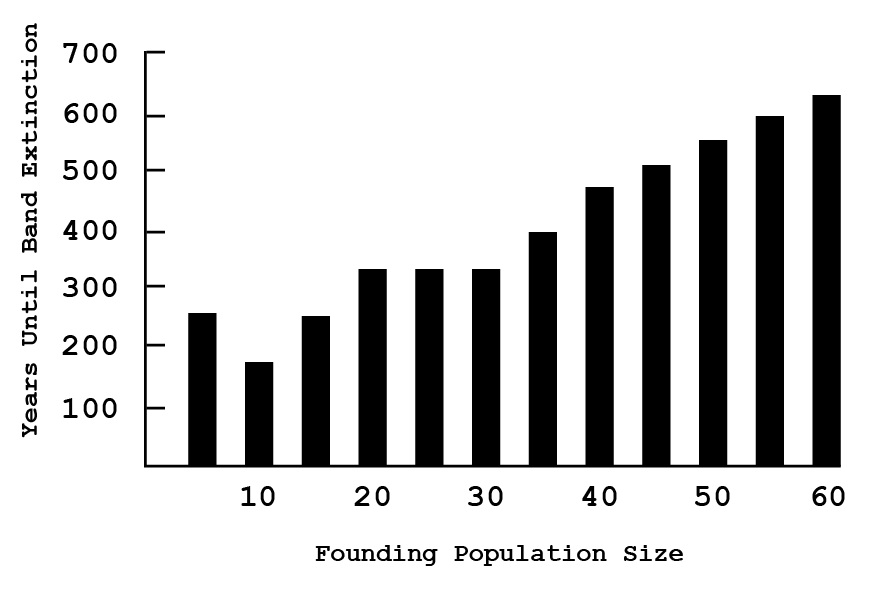}
  \caption{ETHNOPOP simulation showing years to demographic extinction 
	   for closed human populations. Bands of people survive longer 
	   with larger starting sizes, but these closed populations all 
	   eventually become extinct due to demographic (age \& sex structure) 
	   deficiencies. Figure adapted from \cite{Moore2003}.}
  \label{Fig7}
\end{figure*}
\begin{figure*}[!t]
  \centering
  \includegraphics[width=0.8\linewidth]{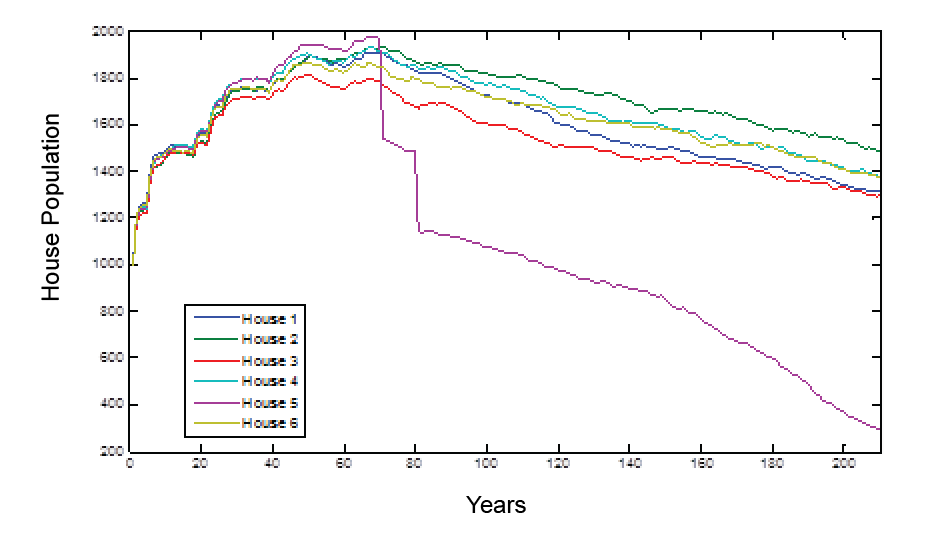}
  \caption{MATLAB results simulating independent population fates over 210 years. 
	  Each ``house'' is a separate population, for example, a village or 
	  largely-isolated or isolable (for quarantine purposes) interstellar 
	  voyaging ‘world ship' traveling, perhaps, in parallel. Each population 
	  begins at 1,000 individuals and is allowed to double in about three generations 
	  to grow to just under 6,000, the pre-set carrying capacity. The collapse 
	  of House 5 came in the form of an ‘extinction vortex', discussed in the text, 
	  in which a random catastrophe so depopulated the breeding population that 
	  age- and sex- structures were disturbed, making it increasingly difficult 
	  to find mates. The general decline in populations resulted from aging 
	  populations and concomitant low replacement levels.}
  \label{Fig8}
\end{figure*}
\begin{figure*}[!t]
  \centering
  \includegraphics[width=1\linewidth]{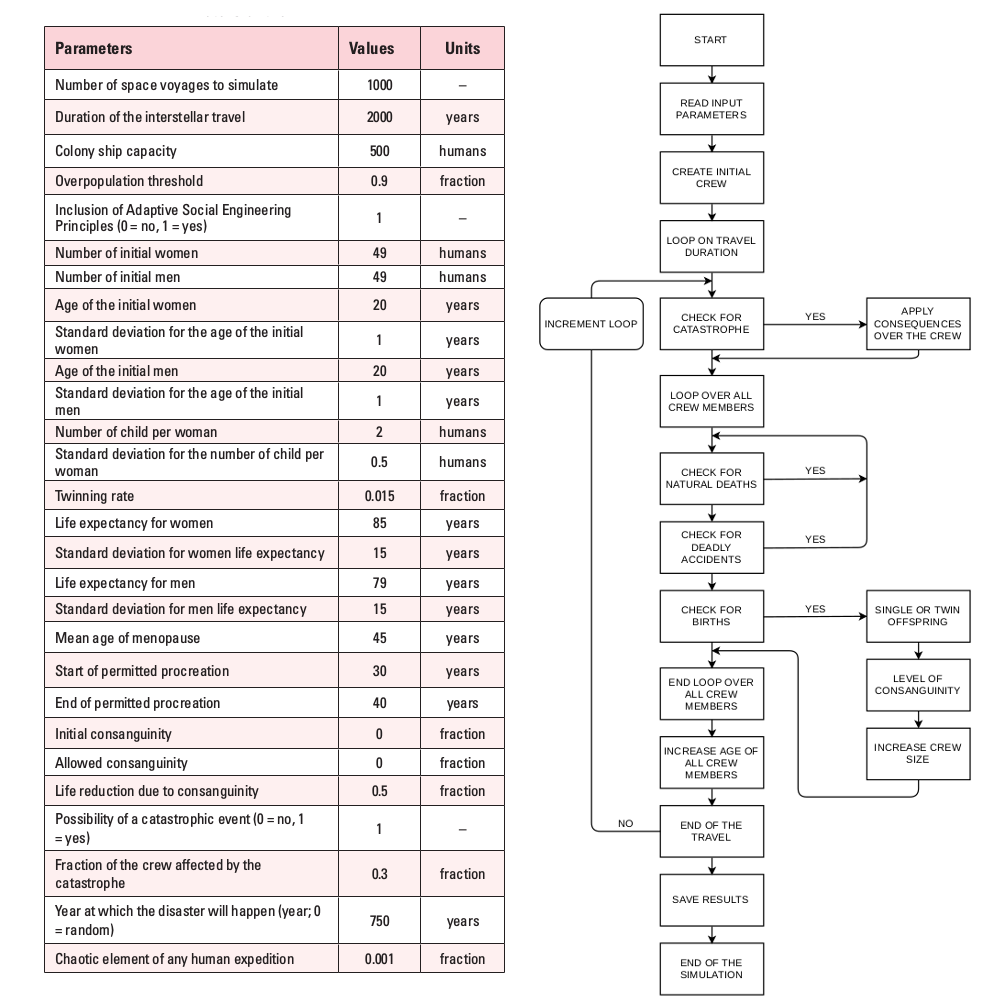}
  \caption{HERITAGE input parameters and flow diagram.}
  \label{Fig9}
\end{figure*}

\begin{figure*}[t]
  \centering
  \includegraphics[width=0.8\linewidth]{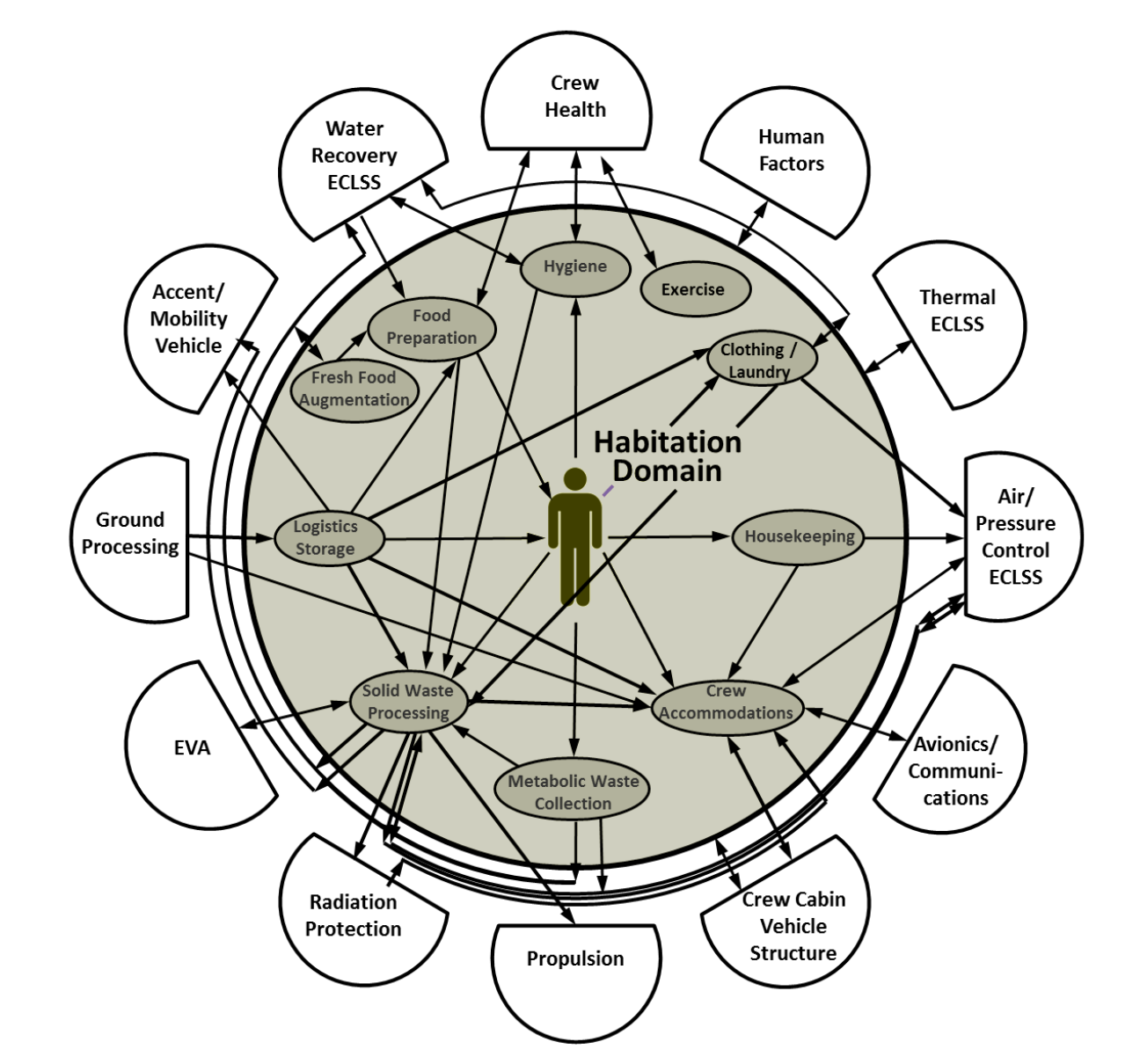}
  \caption{Selected model agents and interactions in SIMOC. 
	  Planning for permanent space settlements will be 
	  radically improved by using dynamic models of variables, 
	  interactions and both expected and emergent properties. 
	  Image courtesy of Kai Staats, MSc.}
  \label{Fig10}
\end{figure*}
How many humans are required, as a founding population, to ensure that future generations live in good multi-generational health? 
The question has been addressed mainly by (a) population geneticists, for theoretical interest, (b) conservation biologists, to help
conserve species at risk of extinction, (c) anthropologists, with an interest in human mating patterns and `prehistoric dispersals, 
(d) space-settlement planners envisioning open populations that may be expected to continue to bring in new members over time and (e), 
space-settlement planners envisioning closed populations that cannot be expected to acquire new members over the course of some journey. 
Our world ship interest is in the latter category.\\
\indent Current world ship plans do not suggest the world ship to return to Earth or to voyage indefinitely. While endlessly-voyaging interstellar vessels might be constructed, in this paper we focus on interstellar voyaging with a definite destination, such as on an exoplanet or free-floating space colonies constructed in a star system. This population is 
envisioned as departing Earth, traveling to its destination, and then establishing the population in a way that it may grow. Such a voyage then includes two founding populations, the founders departing Earth and the founders serving as the original stock on 
the exoplanet from which subsequent generations will derive. Establishing the first founding population is envisioned to be relatively 
easy, as humanity will have the Earth's diversity and population of people from which to choose a genetically healthy gene pool, 
composing a population. A recent review of the possibilities of genetic and cultural ‘screening’ for world ship populations has found that there is no need to envision such a process; there is no conceivable reason that people of just about any genetic composition, including genetic disorders, could not compose some portion of the voyaging population, and, as on Earth, live out happy and productive lives \cite{Smith}. Such ‘screening’ is particularly difficult to justify when starting to explore a literal universe of unknown selective pressures. The constitution of the second founding population, however, is of most interest as it must itself be intact 
and healthy after the period of travel. The question then becomes, what do we need to do to ensure a bio-culturally healthy population 
after some centuries as a closed population? Note that this is a more involved question than ``what is the minimum number of people 
who can survive a multi-century voyage to an exoplanet''. We return to this issue below after a review of the main estimates so far 
made to inform the world ship-planning community. Seven carefully-researched estimates of interstellar voyage founding populations have 
been published, see Table ~\ref{Tab2}. For reference here we will distinguish between the founding population departing Earth (D1 for deme 1) 
and founding population arriving at an exoplanet (D2 for deme 2). Table ~\ref{Tab2} indicates these five estimations and also SIMOC, an 
agent-based model currently in development.\\
\indent Anthropologist John Moore (1935 -- 2016) referenced ETHNOPOP, a program written in C++ with his colleagues Dan Yu and Wenqiu Zhang 
in the 1990's, in his 2003 book chapter ``Kin-based Crews for Interstellar Multi-Generational Space Travel'' \cite{Moore2003}. His 
computer program was based on his extensive knowledge of hunter-gatherer band and tribe mating practices that linked small bands 
of travelers genetically over large landscapes. Essentially Moore's model was concerned with demographics, the age- and sex-structure
of a given population, because of his original interest in using it to model prehistoric human dispersals across the globe. The
program began with a population of a given size, and age- and sex-structure, and at each computing cycle (representing a generation, 
or circa 30 years), evaluated each population member's likelihood of mating, death, and a few other variables. Thus it was an agent-based 
model, though of limited power compared to today's models. Still, his application of the model to spacefaring suggested to him that a 
founding population (D1) of 80 -- 150 individuals would be sufficient to avoid inbreeding over a multi-century voyage (see Fig.~\ref{Fig7}). 
The second founding population (D2) mentioned above was not a concern for Moore as he actually proposed that his voyagers would return 
to Earth, where they could again have the luxury of mating into a large and diverse population.\\
\indent Subsequently in 2012 Cameron Smith took on the task of identifying an ``Interstellar Migrant Population'' (D1) for Icarus Interstellar's 
Project Hyperion, to assist in the reference design for a world ship. The approach, published in 2014 as a research paper titled ``Estimation
of a Genetically Viable Population for Multigenerational Interstellar Voyaging: Review and Data for Project Hyperion'', was statistical 
and largely based on population genetics. Smith surveyed the research literature for various animal species' Minimum Viable Populations,
figures below which natural animal populations did not drop in nature. With these as a context, he considered humanity's bio-culturally-evolved
natural populations, which, in the circa 10,000 range, were not so different from the average for vertebrates. He considered also the effects of 
the primary population-dynamics processes of genetic mutation, migration, selection and drift, as well as the likelihood of catastrophe en 
route to a given exoplanet. From these figures a simple formula to estimate the ideal D1 population given the desired D2 population was derived,
suggesting ``anywhere from roughly 14,000 to 44,000'' individuals as entirely safe D1 populations sufficient to ensure a D2 population equal
to or greater than the common human breeding population of c.10,000 \cite{Smith2014b}. Lower figures for D1, in the 7,500 range, Smith 
mentioned, were on the edge of reasonable, but he strongly suggested that 10,000 should be the absolute minimum D1 population setting off 
from Earth. His paper's final approach was to run agent-based computer simulations, written in MATLAB by his colleague William Gardner-O'Kearny. 
These were agent-based, establishing a D1 population and observing its change over each computing cycle, during which —-- as in Moore's 
model —-- each agent's age, probability of death, probability of finding a mate (based on its own age and sex), and other variables, was computed, 
such that the population changed over time. This model revealed that, strictly demographically, populations in the circa 2,000 range could survive 
for some centuries as closed systems (see Fig.~\ref{Fig8}). While noting this, I did not suggest it as a viable D1 population, for reasons we 
will return to below.\\
\indent In 2017 astrophysicist Fr\'ed\'eric Marin published the first in a series of papers (ongoing) revealing results of his HERITAGE agent-based 
program titled ``HERITAGE: A Monte Carlo Code to Evaluate the Viability of Interstellar Travels Using a Multigenerational Crew'' \cite{Marin2017}. 
The approach used Monte Carlo methods (repeated random sampling of repeated simulations) to identify system properties emergent over time due 
to the properties of many individual agents. HERITAGE was written in C/C++ and is characterized by a large number of variables being 
evaluated at each computing cycle per individual agent. Access to superior computing power allowed Marin to run many hundreds or thousands 
of these simulations in the Monte Carlo approach (in the U.S. a related method is referred to as ‘bootstrapping') to identify general system 
properties based on many specific cases (simulated voyages). Fig.~\ref{Fig9} indicates the HERITAGE flowchart and input parameters. Marin and his 
colleagues ran multiple simulated multi-century voyages that include the following results. Simulations of Moore-like populations 
(D1 population set at 150) over 200 years resulted in unhealthy inbreeding (due to the small population) and population reduction at the 
end of the voyage to about 33\% of the original population, so that D2 (exoplanet founder population) would be about 50 individuals. 
Simulations of populations based on Smith figures of D1 set at 14,000 people departing Earth were shown to ``be more efficient at mixing the 
genetic pool in order to ensure a safe sixth generation [D2 founder population] ... and even with a severe catastrophe the mission is not 
compromised ... [this scenario] is the only one to achieve the goal of ... bringing a genetically healthy crew to another distant planet'' \cite{Marin2017}.
Marin's second HERITAGE paper (``Computing the Minimal Crew for a Multi-Generational Space Travel Towards Proxima Centauri~$b$''), 
coauthored in 2018 with particle physicist C. Beluffi, applied an improved version of HERITAGE which included more complex mating and other 
reproductive rules, again evaluated at each computing cycle for each simulated member of the population \cite{Marin2018a}. In this simulation, 
populations starting with D1 numbering 150 individuals (and a world ship capacity of 500) survived not only centuries, but over six millennia, 
in good genetic and demographic health; an even smaller D1 figure (98 individuals) was also identified as viable on this timescale. This was
attributed  to ``adaptive social engineering principles'' that change the mating rules en route, rather than applying a single rule throughout the 
voyage. This is an entirely reasonable adjustment of HERITAGE and is encoded in IF-THEN constructs such as the following: ``If the amount of 
people inside the vessel is lower than the [world ship's predetermined capacity], the code allows for a smooth increase of the population by 
allowing each woman to have an average of 3 children (with a standard deviation of 1). When the threshold is reached, HERITAGE impedes the couples' 
ability to procreate but allows women that were already pregnant to give birth even if the total number of crew members becomes marginally higher 
than the threshold.'' \cite{Marin2018a} While the study proposed very low D1 figures (compared to any other estimates) the authors did caution in
the paper that ``the impact of mutation, migration, selection and drift is not included in HERITAGE ... [so] we emphasize that the minimum crew 
of 98 settlers we found is a lower limit ...'' and that further work might well suggest a larger D1 figure. The main advantage of HERITAGE is 
that it more accurately models real human mating behavior, which is not random and can thus, by consciously avoiding inbreeding, support smaller 
populations. In Marin's third paper reporting results of HERITAGE (``Numerical constraints on the size of generation ships from total energy 
expenditure on board, annual food production and space farming techniques''), 500-person D1 populations were used as reference study
\cite{Marin2018a}. The authors addressed the crucial question of how to feed the crew, since dried food stocks are not a viable option due 
to the deterioration of vitamins with time. The best option then relies on farming aboard the spaceship. Using an updated version of HERITAGE, 
Marin's team were able to predict the size of artificial land to be allocated in the vessel for agricultural purposes. \\
\indent Although no results have so far been published, the SIMOC (Scalable Model of an Isolated, Off-World Community) multi-agent simulation is 
near completion. Orchestrated by Kai Staats as a project of the Arizona State University School of Earth and Space Exploration's 
Interplanetary Initiative, SIMOC is designed to model and then analyze the results of the physical characteristics of an off-Earth 
colony. In particular the habitat's agricultural, life-support, recyclable and consumable variables are modeled, as is the health of each 
colonist placed in the system (see https://simoc.space/): ``to design a habitat that sustains human life through a combination of 
physio-chemical (machine) and bio-regenerative (plant) systems, and then scales over time, with SIMOC Phase IV -- V including options to grow 
the community with the addition of inhabitants and infrastructure ... [based on] ... an agent-based model (ABM), a class of computational models 
for simulating the actions and interactions of autonomous agents (both individual or collective entities such as organizations or groups) 
with a view to assessing their effects on the system as a whole.'' \cite{Smith}.\\
\indent The project's developmental phases are described below; at this writing the project is in Phase IV -- V with expected activation and public 
release in the first quarter of 2019:
\begin{itemize}  
\item Phase I: Habitat modeling: low-Earth orbit, on the Moon, in free space, or on Mars. Attention was given to specific locations, 
such as a valley, mountain top, or polar cap as each would inherit a particular in situ resource utilization (ISRU) parameter;
\item Phase II: Physio-chemical modeling of ECLSS (Environmental Control and Life Support System) and bio-regenerative systems;
\item Phase III: Agent modeling \& integration with Phase I and II module;
\item Phase IV: Population modeling. Consumables tracking ; modeling which construction materials shipped from Earth versus were 
manufactured locally, via ISRU (In-Situ Resource Utilization); each expansion task is restricted by the cost of energy and time;
\item Phase V: Modeling aging of the systems and stochastic (entropic) breakdown such as habitat gas leaks, solid waste processor 
failures, or a space bolide strike resulting in catastrophic failure of a greenhouse and all crops therein.
\end{itemize}
SIMOC is currently configured to model, as mentioned, physical rather than social dynamics, although the designers have expressed an 
interest in the interactions and emergence of social phenomena (see Fig.~\ref{Fig10} for a summary of the potential SIMOC agents and
interactions). Such social phenomena have been addressed in the field of multi-agent social simulations, capably defined as ``... the 
intersection of three scientific fields, namely, agent-based computing, the social sciences, and computer simulation ...'' \cite{Davidsson2002}.
In the future, it will be very interesting to compare the results of SIMOC and HERITAGE. Current plans include comparison of SIMOC simulations
with real-world closed habitat experimentation at the University of Arizona, a form of ``ground truthing'' in which a mathematical model 
may be improved by comparison with, and then better modeling of, real-world systems. Social phenomena that may emerge in SIMOC and other
multi-agent simulations could be of great interest. At this writing we are aware of, but have not been able to review, W.S. 
Bainbridge's 2019 book, ``Computer Simulations of Space Societies'' \cite{Bainbridge2018}. 

\subsection{Reasons for estimate variations for D1, Earth-departing interstellar population}
It is natural that a variety of population sizes have been proposed for D1, the Earth-departing founding population, as researchers from 
different backgrounds have brought various approaches to this question. We believe that some of the variation derives from different 
conceptions of human populations and human behavior over time.\\
\indent Moore's gravitation towards low figures come from a long-term anthropological perspective recognizing that hunting and gathering cultures 
have survived for many thousands of years in low population densities, so that just few centuries should be relatively easy for a D1 
population less than several hundred. However, Moore's figures appear somewhat too low as he did not really account for the fact that while 
humans may live together in breeding populations (demes) numbering in the hundreds (the famous ``Dunbar Number'' of about 150 individuals is often 
quoted regarding hunter-gatherer group size \cite{Dunbar1993}), such populations always have reproductive links with other groups. Also, 
his figures largely reflect populations of hunting-and-gathering folk who move seasonally over large landscapes, whereas in all conceivable
world ship designs the subsistence mode would be agriculture, which is characterized by residential sedentism. However, the type of agriculture might be diverse, including hydroponic and aeroponic farming, potentially extending to the use of emerging technologies such as artificial meat \cite{Marin2018}. Residential sedentism, 
worldwide and throughout prehistory, always leads to higher populations, as we introduce below. \\
\indent Smith's anthropological biases led him highlight larger population figures because human populations are always linked to others, normally 
in the thousands of individuals, figures approaching the circa 7,500 population range for naturally-evolved populations of naturally-occurring 
vertebrates. He is also conditioned by an emphasis on catastrophe: for Moore, human populations have generally survived quite well even small 
populations in particular because they have often had large landscapes and many resources available; a local catastrophe could be averted by 
moving to new resource territories, and if one group actually became extinct, humanity was so widespread that others always continued. But 
for Smith, considering the perspective of a closed population carrying all their resources with them, there is an expectation that eventually 
some catastrophe will strike, and for this single, isolated population there is nowhere to go, no ``geographical reserve''. To be sure to arrive 
in relatively safe populations (D2), Smith has gravitated towards particularly large departing populations (D1).\\
\indent For Marin's approach with C. Beluffi, there is an attempt to reduce D1 to an absolute minimum as revealed in the paper title, with the paper's 
function stated as ``to quantify the minimal initial crew necessary for a multi-generational space journey to reach Proxima Centauri~$b$ with 
genetically healthy settlers''. Here the focus is on propulsion at speeds achieved today with the Parker Solar Probe, resulting in the need 
to voyage for an estimated 6,300 years to Proxima Centauri~$b$. The philosophy driving the search for the minimum viable population here is
that of the ``scarcity paradigm'' of crewed spaceflight. In this paradigm, we must identify the minimum mass to transport to reduce cost. 
Marin's comparatively small figure of less than a hundred individuals is identified as a viable D1 figure under very strict adaptive mating rules 
that may change over the course of a journey back, then, to figures closer to the Moore thought-scape.

\subsection{Biological health: where are we today on estimates of world ship populations?}
From strictly mathematical, statistical and genetic perspectives we may say that Earth-departing D1 founder populations of humans, numbering 
in just the low hundreds of people, could theoretically survive for centuries or even some millennia in health sufficient to serve as D2 
(exoplanet founder) populations when mating is cleverly devised to avoid inbreeding. Smith mentioned this in the 2014 paper, for instance, 
stating that ``any population over 100 or so'' would avoid some of the chief problems of small populations on such timescales. Marin 
demonstrated this with the high-fidelity HERITAGE program that capably simulates human social engineering to manage population health.\\
\indent While the smallest figures may work biologically, 
they are rather precarious for some generations before the population has been allowed to grow. We therefore currently suggest 
figures with Earth-departing (D1) figures on the order of 1,000 persons. Because useful modeling is still 
underway, there is a practical way to use such an estimate even at this early date. We propose that for habitat design and modeling, 1,000-person
modules (alternatively called villages) be designed, that could at a later date simply be multiplied as elements of a world ship cluster. This 
way, the Earth-departing population could be set to any figure one wants, for example 3 villages composing 3,000 people, or 10 for 10,000 or 
just one for a departing population of 1,000. While this modularity does increase mass (as compared to a single-vessel design using the most 
efficient enclosure of space by material) and thus the budget to be allocated for such large missions, we feel the modularity is worth the 
trade-off. For instance, multiple, independent villages traveling in parallel, each with a population of circa 1,000, would reduce the possibility
of a catastrophe wiping out the entire population. The ‘villages’ would travel together on the same spacecraft but would be somewhat separated, with the possibility to allow travel from village to village. Traveling in parallel would allow people to visit other ``towns'' for pleasure, cultural exchange and marriage and reproduction, but also to be quarantined (culturally and/or biologically) if desired. Such a concept of interacting habitats was previously proposed by Sherwood for space colonies within our Solar System \cite{Sherwood1989}. The population on the order of 
1,000 per village module is also viable culturally, as we explore below. A more in-depth analysis of this topic is provided in \cite{Smith}. 

\subsection{Features of successful world ship population cultures}

\begin{figure*}[t]
  \centering
  \includegraphics[width=0.6\linewidth]{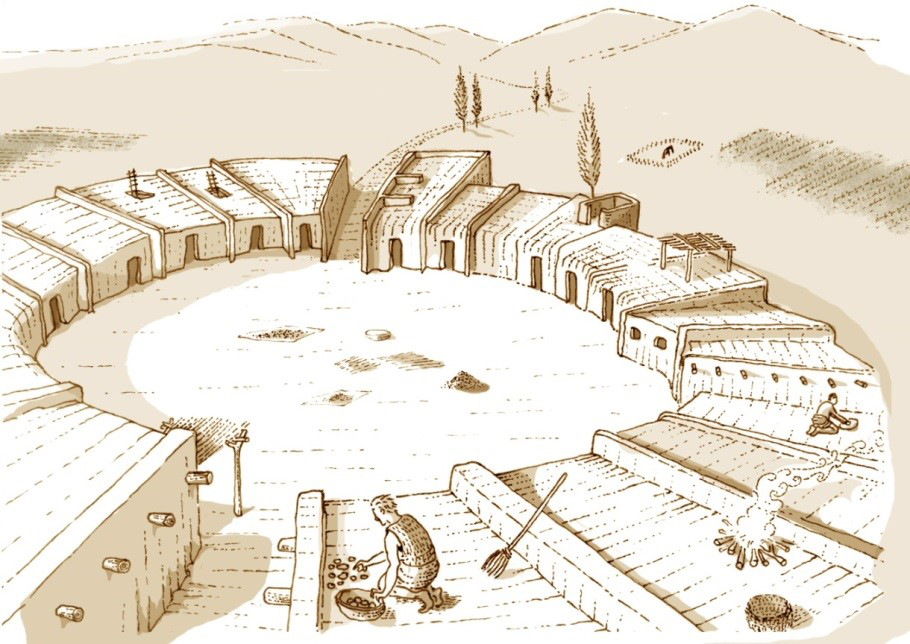}
  \includegraphics[width=0.6\linewidth]{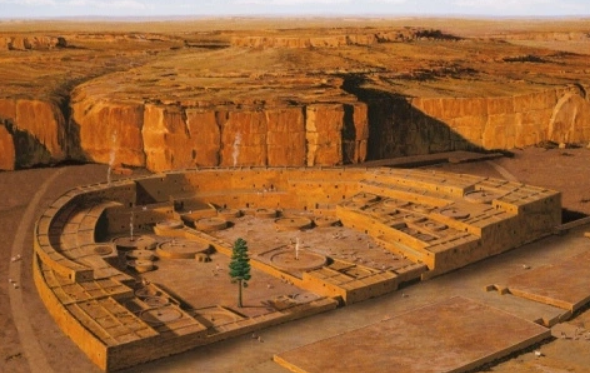}
  \caption{Reconstructions of Two Independent Neolithic Villages. 
	   Top: Demircihuyuk, Turkey (drawn by Cameron Smit); 
	   bottom: Chaco Canyon.}
  \label{Fig11}
\end{figure*}

\begin{table*}[!t]
  \centering
  \caption{Independent neolithic village population estimates. 
	   See text for discussion.}
  \begin{tabular}{p{0.19\linewidth}p{0.25\linewidth}p{0.18\linewidth}p{0.22\linewidth}}
  \hline
  Region & Village & Date (years ago) & Population estimate \\
  \hline
  1. SW Asia & Jericho & 10000 & 225\\ 
  1. SW Asia & Netiv Hagdud & 10000 & 135\\ 
  1. SW Asia & Gilgal I & 10000 & 90\\ 
  1. SW Asia & Dhra' & 10000 & 41\\ 
  1. SW Asia & Nahhal Oren & 10000 & 18\\ 
  1. SW Asia & Ain Ghazal & 8900 & 405\\ 
  1. SW Asia & Tell Aswad & 8900 & 360\\ 
  1. SW Asia & Jericho & 8900 & 225\\ 
  1. SW Asia & Yiftahel & 8900 & 135\\ 
  1. SW Asia & Kfar Hahoresh & 8900 & 45\\ 
  1. SW Asia & Catalhoyk & 8600 & 6000\\ 
  1. SW Asia & Basta & 8250 & 1260\\ 
  1. SW Asia & Ain Ghazal & 8250 & 900\\ 
  1. SW Asia & Wadi Shu'eib & 8250 & 900\\ 
  1. SW Asia & Beisamoun & 8250 & 900\\ 
  1. SW Asia & Es-Sifiya & 8250 & 900\\ 
  1. SW Asia & Ain Jamman & 8250 & 630\\ 
  9. Europe & Cyprus & 6000 & 2000\\ 
  9. Europe & Serbian sites & 6000 & 1740\\ 
  3. East Asia & Xinglongwa & 7730 & 100\\ 
  3. East Asia & Cishan & 7700 & 100\\ 
  3. East Asia & Zhaobaogou & 7034 & 100\\ 
  9. Europe & Germany & 6000 & 135\\ 
  4. Africa & Merimda Beni Salama & 6000 & 1650\\ 
  4. Africa & Hierakonpolis & 5500 & 1750\\ 
  6. South America & Real Alto & 5250 & 175\\ 
  6. South America & Loma Alta & 4680 & 175\\ 
  2. South Asia & Ban Non Wat & 4000 & 700\\ 
  5. Mesoamerica & Oaxaca sites & 3300 & 325\\ 
  5. Mesoamerica & Oaxaca sites & 2900 & 1973\\ 
  5. Mesoamerica & Oaxaca sites & 2770 & 1782\\ 
  5. Mesoamerica & Oaxaca sites & 2600 & 1828\\ 
  5. Mesoamerica & Oaxaca & 2600 & 1000\\ 
  5. Mesoamerica & Basin of Mexico sites & 3050 & 685\\ 
  \hline  
  \end{tabular}
  \label{Tab3a}  
\end{table*}
\begin{table*}[!t]
  \centering
  \caption*{Table ~\ref{Tab3a} continued.}
  \begin{tabular}{p{0.19\linewidth}p{0.25\linewidth}p{0.18\linewidth}p{0.22\linewidth}}
  \hline
  Region & Village & Date (years ago) & Population estimate \\
  \hline
  6. South America & Titicaca basin sites & 3250 & 693\\ 
  6. South America & Titicaca basin sites & 2900 & 1752\\ 
  6. South America & Titicaca basin sites & 2500 & 3507\\   
  8. North America & Snaketown & 1000 & 300\\ 
  8. North America & Galaz & 1000 & 300\\ 
  8. North America & Montezuma Valley & 800 & 2500\\ 
  8. North America & Yellowjacket & 800 & 2000\\ 
  8. North America & Zuni & 800 & 1600\\ 
  8. North America & Sand Canyon & 800 & 725\\ 
  8. North America & Marana & 800 & 700\\ 
  8. North America & Paquime & 600 & 4700\\ 
  8. North America & Sapawe & 600 & 2770\\ 
  8. North America & Pueblo Grande & 600 & 1750\\ 
  8. North America & Los Muertos & 600 & 800\\ 
  7. Amazonia & Rio Negro Sites & 2300 & 1250\\ 
  7. Amazonia & Upper Rio Xingu Sites & 1000 & 1250\\ 
  7. Amazonia & Central Brazil & 1000 & 964\\ 
  8. North America & Chaco Canyon & 1300 & 600\\ 
  8. North America & SW USA & 1300 & 400\\ 
  8. North America & Mesa Verde & 1100 & 100\\ 
  8. North America & Chaco Canyon Main Village & 1000 & 3500\\ 
  8. North America & Chaco Canyon hamlets & 1000 & 200\\ 
  8. North America & Moundville & 1000 & 1200\\ 
  8. North America & Snodgrass & 1000 & 350\\ 
  8. North America & Lunsford-Pulcher & 950 & 1000\\ 
  8. North America & Cahokia & 950 & 1000\\
  \hline
  \textbf{Average} & ~ & 4444 & 1088\\  
  \hline
  \textbf{Standard Deviation} & ~ & ~ & 1163\\  
  Low & ~ & ~ & 18\\ 
  High & ~ & ~ & 6000\\ 
  \hline  
  \end{tabular}
  \label{Tab3b}   
\end{table*}

\begin{figure*}[t]
  \centering
  \includegraphics[width=0.8\linewidth]{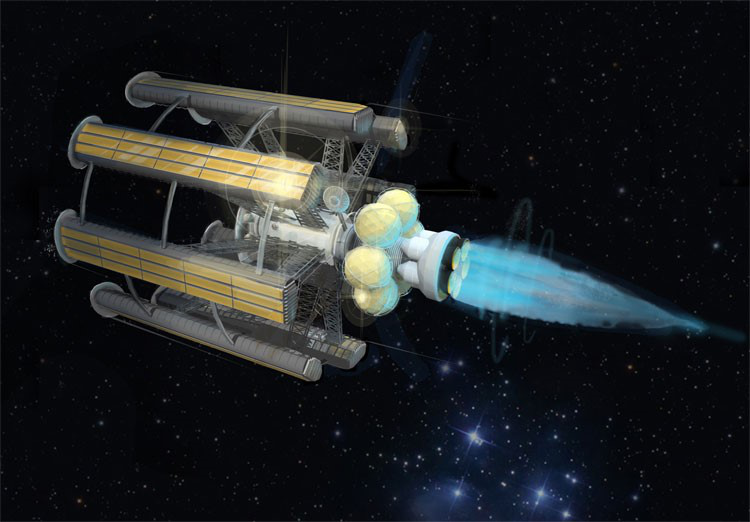}
  \caption{A multi-generational vessel schematic. 
	   Revolving for 1-g conditions around a central hub are eight habitations; 
	   We suggest each to be self-supporting, but allowing communication 
	   with others, and that each could have a population on the order of 
	   500 -- 1,000, much like independent farming villages today and in 
	   the past. Propulsion and other systems are kept at some distance to 
	   the rear. Figure copyright and courtesy of Steve Summerford.}
  \label{Fig12}
\end{figure*}
Before the interconnection of the modern world, and before the radical changes of urbanism that characterize modern and ancient civilizations, 
early farming people worldwide lived in independent farming villages with many features we think will be analogous to those of interstellar 
voyagers. For example, Marin et al. \cite{Marin2018} have used HERITAGE to also model on-board food production, indicating that dried food
stocks are not a viable option due to the deterioration of vitamins with time and the tremendous quantities that would be required for 
long-term storage. Having a sustainable source of food is thus mandatory for such long journeys and the space needed for geoponics (or 
hydroponics/aeroponics) will strongly condition the architecture of the spacecraft. Among other results, Marin et al. found that for an 
heterogeneous crew of 500 people living on an omnivorous, balanced diet, 0.45~km$^2$ of artificial land would suffice in order to grow 
all the necessary food using a combination of aeroponics (for fruits, vegetables, starch, sugar, and oil) and conventional farming (for
meat, fish, dairy, and honey). This translates into various spaceship lengths and radii, depending on the level of artificial gravity we want 
to produce on-board.\\
\indent To learn from humanity's long experience of farming in independent farming villages we note first that those populations were rather 
self-supporting. While there was trade, it was not global, but among multiple villages in a relatively small region. This is much like any 
world ship considered today; certainly trade will be rather local, which in part shapes the economy. These villages were also unfortified; 
while social friction did occur, so much time was spent in food production and processing that it was not possible to maintain standing 
military forces; such is also identifiable in most world ship plans. Early farming villages also had a rather domestic economy, where if 
you needed something, you generally made it yourself. Certainly there were some specialists, but there was a more general self-sufficiency 
of fabrication. On reasonably-expectable world ships we feel something very similar will play out at least in the lack of emphasis on, 
again, a large trade in products. Rather many items will be fabricated locally and on the scale of the household or community rather than 
on the scales of a global market. Early independent farming villages were also horticultural, rather than agricultural. That is, while 
they did farm, the farming was again of a local character, serving communities or households, rather than for a market of millions or 
billions, and again this will be similar in world ships with total populations perhaps less than several tens of thousands. Early farming
villages also had populations in the 600 to 1,000 range, similar to world ship estimates we see above. Fig.~\ref{Fig11} illustrates such 
villages at Demircihuyuk, Turkey, and Chaco Canyon, New Mexico.\\
\indent Table ~\ref{Tab3a} presents summary population estimates of early farming villages, worldwide (data derive from Smith 2019, in press.). 
As mentioned earlier, the village populations were managed in the low thousands, often around 1,000. Villages were some kilometers from
one another, such that while there were interactions with others, such that while there were interactions with others, each village was self-sufficient. Self-sufficiency means here that a local production and consumption system exists.  That such populations
managed as relatively stable and self-sufficient units for some millennia (in many cases for several thousand years before the advent of
civilization) in arrangements that have important similarities to how we imagine interstellar world ships today has caused us to 
investigate them in some detail. For the moment we will simply say that they may be useful analogues for world ship design considerations.  
Fig.~\ref{Fig12} is the original illustration of a multi-community world ship published in \cite{Smith2014}, designed and provided by 
urban designer Steve Summerford. With the insight of the HERITAGE we feel it is safe to reduce the population from 5,000 per each of 
the eight villages (originally proposed) to 500 or 1,000. This would bring the D1 population to about 4,000, organized something like 
the highly-successful early independent farming villages in humanity's collective early experience, and not so small as to be terribly 
vulnerable. Finally, such populations are familiar in the human experience, and we suggest remaining nearer the human experience than 
farther from it, for cultural viability and palatability, especially in a project of such an exotic nature as the world ship voyage.\\
\indent We acknowledge that organizing the population of a world ship into farming villages is not a new idea. Interiors of space colonies and world ships have been regularly depicted as  rural or suburban areas with sparse habitations \cite{ONeill1974,Johnson1977,Bond,Arora2006} with Paolo Soleri's Asteromo as an exception \cite{Soleri1973}. More recent proposals for world ship habitats imagine evolving structures that adapt to the population during its trip \cite{Armstrong}. It is important to point out that we are not prescribing any particular interior design. These might be designed by the world ship-farers and builders themselves. \\

\subsection{Productive New Ways To Think of Interstellar Voyaging}

\begin{figure*}[!t]
  \centering
  \includegraphics[width=0.94\linewidth]{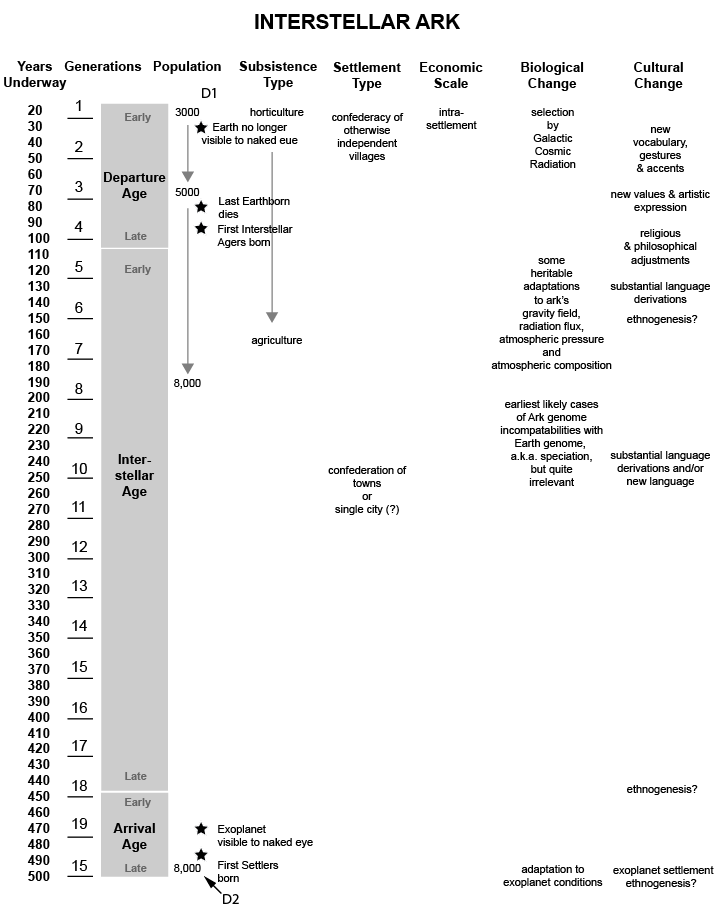}
  \caption{Some hypothetical, but reasonably expected, events of 
	  a multi-generational voyage to an exoplanet. Issues introduced 
	  here are only to sketch out the thought-scape, and they are 
	  discussed further in Smith 2019 (in press).}
  \label{Fig13}
\end{figure*}

What aspects of culture and biology may we productively address with the objective of making the interstellar voyaging project most likely to succeed? Smith \cite{Smith} investigates this question, concluding that we should focus on humanity’s adaptive tools, both biological and cultural. Culturally these include a set of human universals, domains of behavior seen in all cultures that are often adjusted to accommodate new conditions. For example, all human cultures have some conception of a family, a cohabiting unit related often by kinship and cooperating often in resource acquisition. Adaptation of the size and structure of the family to the conditions is clear and many times predictable. For example, foraging cultures tend to have smaller families that can travel nimbly, whereas farming cultures tend to have larger families for the many simultaneous tasks of farming). In this case, the human universal of family size and structure may be investigated for its adaptive range and potential, and how it may be configured for interstellar voyaging conditions. Such an investigation is presented in \cite{Smith} and is too extensive to review here, but the point is that there exist good theoretical reasons to delineate the discussion of culture aboard world ships along the lines of human universals. As a direct consequence, while each world ship might exhibit unique features of its population, they will likely have common features which are a consequence of human universals.\\
\indent We suggest a few anthropologically-guided suggestions that may help to shape more realistic world ship studies. First, we think 
we should move away from the paradigm of scarcity, and towards a paradigm of plenty. Certainly if setting off for a multiple-century
or -millenium voyage, one would wish to travel with a large margin and surplus, not in arrangements that would be just mathematically
possible. Second, we would think about families and communities rather than crews. Crews eventually go home and have a concept 
of home being somewhere else; but on world ships, many will be born who will have no experience of losing Earth or gaining
the exoplanet, they will live out normal, small-town lives in the world ship. Third, we would suggest moving away from thinking 
of mating or reproduction rule as something of a problem for the inhabitants. Indeed we think the people who choose to voyage on 
these vessels will be the folk who construct them in the first place, and they will naturally have rules about reproduction to 
keep their population from exceeding the world ship capacity, just as populations today have plenty of cultural regulations of  various behaviors. Fourth, we would move away from conception of the world ship as a vessel on a mission; again, it will be the home of people who grow, live and die and it is hard to imagine that they will think of themselves on a ship or on a mission 
(except for the earliest and latest generations aboard), rather people will be living normal lives. Finally, we would attempt to 
de-exoticize the interstellar voyage. Fig.~\ref{Fig13} presents some expectable changes we may see in world ship population biology 
and culture over the centuries (or more) of an exoplanet voyage. Time may be divided into departing, interstellar and arriving 
ages; the population may grow (if permitted); the language and biology will change subtly. All of this will be carried out, 
however, on the individual timescales and experiences of normal people living out daily life. It is this anthropological 
perspective that continues to influence our thinking about world ships. 

\section{Socio-technical feasibility}

\begin{figure*}[t]
  \centering
  \includegraphics[width=1.0\linewidth]{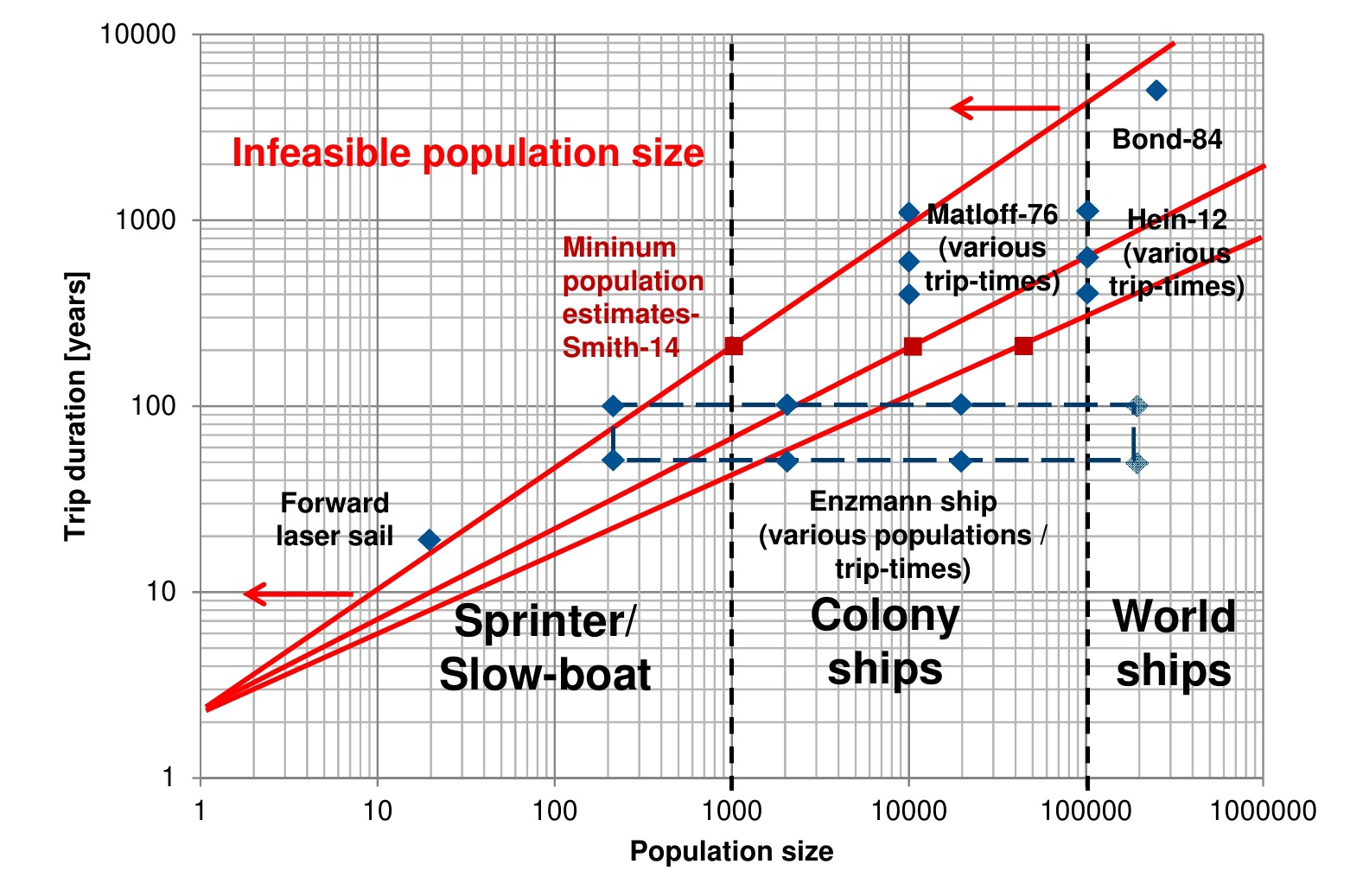}
  \caption{Crewed starship categories versus population size and trip duration.}
  \label{Fig2}
\end{figure*}

World ship feasibility also depends on social and technical factors. In the following, we will present various world ship destinations and what implications this would have for a world ship mission and the settlement activities for developing a new civilization. Subsequently, we present the population - trip duration trade-off, which helps determine which types of missions are feasible. Finally we briefly present previous results regarding world ship reliability.  

\subsection{World ship destinations} \label{S61}

\begin{table*}[!t]
  \centering
  \caption{Potential destinations for world ships.}
  \begin{tabular}{p{0.14\linewidth}p{0.14\linewidth}p{0.14\linewidth}p{0.14\linewidth}p{0.14\linewidth}p{0.14\linewidth}}
  \hline
  ~ & \textbf{Habitable planet/moon} & \textbf{Bio-compatible planet/moon} & \textbf{Easily terraformable planet/moon} & \textbf{Rogue planet/comet} & \textbf{Space colonies} \\
  \hline
  \textbf{Investment for habitability} & Small & Establish ecosystem & Terraforming & Colony construction & Colony construction \\ 
  \textbf{Duration until habitability} & Years & Decade / centuries  & Centuries & Decades & Decades\\ 
  \textbf{Habitability duration} & Millions of years  & Hundred thousands of years & Hundred thousands of years & Centuries -- millennia & Centuries -- millennia\\ 
  \textbf{Availability} & Rare & Rare & Rare & High abundance & High abundance\\ 
  \textbf{Distance from Earth (estimates)} & 4-16 ly & 4-16 ly & 4-11 ly & $\le$ 4 ly & 4 ly\\ 
  \hline  
  \end{tabular}
  \label{Tab4}  
\end{table*}

World ship design is driven by trip time, as mentioned in the Section \ref{S1}. Trip duration, however, is determined by the velocity of the spacecraft and the distance it travels. Distance is determined by the destination to which 
the world ship aims to travel.\\
\indent Since the World Ship Symposium in 2011, a range of new discoveries have been made, which may 
change significantly the range of destinations to which a world ship could travel. \\
\indent In Hein et al. \cite{Hein2012b}, four types of habitats are adopted from \cite{Fogg1991}: habitable planet, bio-compatible planet, easily terraformable planet, and using other resources for constructing free-floating space 
colonies. \cite{Hein2012b} extend the list by adding ``moon'' to ``planet'', due to the potential habitability of exomoons 
\cite{Heller2013,Kaltenegger2010}. Furthermore, so-called rogue planets, which are not bounded to a star and free floating have 
been confirmed via micro-lensing in 2011 \cite{Delorme2012}. Rogue planets could be another type of destination for world ships. A summary of these destinations is given in the following: 
\begin{itemize}
 \item \textit{Habitable planet / moon:} An environment "sufficiently similar to that of the Earth as to allow comfortable and free human habitation." \cite{Fogg1991}
 \item \textit{Bio-compatible planet / moon:} Possesses "the necessary physical parameters for life to flourish on its surface." \cite{Fogg1991}
 \item \textit{Easily terraformable planet / moon:} Can be converted into a bio-compatible or habitable planet with moderate resources available to "a starship or robot pre-cursor mission." \cite{Fogg1991}
 \item \textit{Rogue planet/comet:} Probably similar environment to outer Solar System planets, moons, and minor bodies. 
 \item  \textit{Free-floating space colonies:} Using other resources for constructing free-floating space 
colonies.
\end{itemize}
To our knowledge, rogue planets have not yet been treated as potential destinations for interstellar spacecraft. Due to the limitations of the observational technique of micro-lensing, Jupiter-sized rogue planets have been confirmed at the 
moment. Some of these discovered rogue planets might be brown or red dwarfs. One key criteria for colonization is the existence 
of an in-situ energy source. Rogue planets seem to generate little to no heat and as they are free-floating, there is no star in its vicinity to provide energy. One possible energy source could be fusion fuel such as Deuterium and Helium-3, as in gas
giants in our solar system \cite{Hein2010a}. Therefore, we can imagine several colonization modes for a gas giant rogue planet. 
Either a free floating colony is constructed, possibly by converting the world ship, or colonies could be established in the 
atmosphere of the rogue planet, for example, via balloons \cite{Bond1978}. The atmosphere would be mined using techniques described in \cite{Bond1978}
and \cite{Hein2010a}. In case the rogue planet is a rocky planet, surface or subsurface colonies could be constructed and Deuterium 
mined from water, which is hypothesized to be available under certain conditions \cite{Abbot2011}. However, rogue planets could also serve as an intermediate fueling stop for world ships. This option would only be interesting if the rogue planet could provide resources beyond fuel that justify a deceleration and acceleration of the world ship.\\
\indent Nearby rogue planets are, 
for example WISE 0855-0714 at a distance of 7.27 light years \cite{Luhman2016}. However, it seems likely that rogue planets at a closer distance will be discovered in the future. \\
\indent We expect that colonies on or around rogue 
planets have about the same characteristics as free-floating colonies or on planetary/moon surface colonies. The only 
potential difference is the distance to a rogue planet, which might be much closer than the next star, rendering it easier to reach with a world ship. \\
\indent An updated table of potential colonization destinations from \cite{Hein2012b} can be found in Table ~\ref{Tab4}. In particular, we have updated the distance from Earth for most destinations in light of the latest exoplanet discoveries. Six potentially habitable exoplanets have been discovered within a distance of 16 ly (Proxima Centauri b, Ross 128 b, Tau Ceti e, Luyten b, Wolf 1061 c, Gliese 832 c). It is currently unclear how far these exoplanets fall into the habitable / bio-compatible category. For example, \cite{Howard2018} argue that the intense flares generated by Proxima Centauri would render Proxima Centauri b inhospitable for surface life. \\
\indent Regarding easily terraformable planets/moons, we argue that there are likely such planets/moons existing within 11 ly. The three exoplanets within 11 ly (Proxima Centauri b, Ross 128 b, Tau Ceti e) in principle seem to be suitable for terraforming. For example, Ross 128 b is located in the habitable zone and no obvious showstoppers such as flares from its host star have been detected so far.\\
\indent Habitable planets and moons with some form of biosphere might be a mixed blessing. Such a biosphere might on the one hand reduce the efforts of building a surface colony, as the atmosphere might be (partly) usable. However, as Davies \cite{Davies2013} has pointed out, it is very likely that such a biosphere is incompatible with terrestrial life forms. In such a case, either the life forms imported to the alien biosphere would need to be made compatible, or the two need to be carefully separated. \\
\indent As previously elaborated in \cite{Hein2012b}, the type of destination has implications for the difficulty of the world ship mission. Depending on the destination, building an initial settlement and ultimately establishing a civilization takes more or less time. Also, the risk of failure in doing so is very different. For example, we currently do not know how difficult it is to co-exist on a habitable planet with an existing biosphere. Also, terraforming is likely to be a very risky endeavor, where failure could mean that the planet or moon is rendered permanently uninhabitable. For a more detailed discussion, see \cite{Hein2012b}. 

\subsection{Population - trip duration trade-off}

As demonstrated in Section \ref{S5}, estimates for required population sizes correlate with trip duration. The longer the trip duration, the higher the required population size. In Fig.~\ref{Fig2}, we show population size and trip duration for various crewed interstellar spacecraft concepts in the literature, using the population estimates from \cite{Smith2014b}, with the discussion presented in this paper. The lower and upper estimates are represented as red squares for a trip duration of 210 years. The three red lines represent an interpolation between population size values for short-term missions (Mars mission with a crew of 3-6 and duration of 2-3 years) and the estimates from \cite{Smith2014b}. The area left of the red line is considered infeasible from a population size perspective. Hence, this chart can be used to evaluate whether or not a world ship design is feasible from a trip duration - population perspective. Furthermore, it allows for making trade-offs between trip duration, which is linked to velocity and energy, and population size, which is linked to spacecraft mass. For example, world ship designers may choose a slower but larger world ship with more people on board. Or they may choose a faster world ship with a smaller population. In any case, they would need to ensure that they are on the right side of the red line. For minimizing risk, they are likely to add a margin to the red line to be on the safe side. \\
\indent Several world ship designs from the literature are put into the chart, such as Matloff-76 \cite{Matloff1976}, Bond-84 \cite{Bond}, Hein-12 \cite{Hein2012b}, and the Enzmann ship \cite{Crowl2012a}. In case several values were given in the reference, such in the case for Matloff-76, Hein-12, and the Enzmann ship, they were also represented in the chart. In particular for the Enzmann ship, the population size does not stay constant but increases 10 times during the trip, which leads to the dashed-line square with two population values for one Enzmann ship concept and two trip durations. The chart shows that the upper estimates for population values from \cite{Smith2014b} would render most of the world ship designs infeasible, except for the Enzman world ship design. For making the infeasible designs feasible, either trip times would need to be decreased or population size increased. \\
\indent As a side note, We have added Robert Forward's crewed laser sail starship from \cite{Forward1984}, which would fall under the category of ``sprinter''. 

\subsection{Reliability}
World ship reliability is likely to be a major feasibility issue, due to the large number of parts and the long mission duration \cite{Hein2012b}. As \cite{Gitelson2003} remarks, the mechanical and electronic components of a bioregenerative life supporting system are much more likely to fail than its biological components. Previously, \cite{Hein2012b} developed a reliability model for world ships. They demonstrate that reasonably high reliability values are only possible if components are either replaced by spare parts or replaced by repaired parts. The number of components that need to be replaced ranges from three per second for a 99.99\% reliability value to one every 20 seconds for a 85\% reliability value, as shown in Table \ref{Tab7}. 

\begin{table}[b]
  \centering
  \caption{Component replacement rates for world ship reliability values \cite{Hein2012b}}
  \begin{tabular}{p{0.3\linewidth}p{0.5\linewidth}}
  \hline
  \textbf{Reliability} & \textbf{Replacement rate [1/s]}\\
  \hline
  99.99\% & 3 \\ 
  85\% & 0.05 \\ 
  \hline  
  \end{tabular}
  \label{Tab7}  
\end{table}

Detecting, replacing, and repairing components at these rates seem to be infeasible for the crew. \cite{Hein2012b} therefore conclude that an automated system is needed. Furthermore, world ship components need to be easily accessible and modular, in order to facilitate replacement. Nevertheless, given the complexity of a world ship, the maintenance system likely needs to be very sophisticated and requires an advanced artificial intelligence such as for the Daedalus probe \cite{Bond1978} or probes described in \cite{Hein2018f}. \\
\indent One way to address world ship reliability could be the substitution of mechanical, electronic, and software components by deliberately engineered biological components, which exhibit self-healing capabilities \cite{Armstrong}. This might also work the other way around. Mechanical, electronic, and software components could exhibit self-healing capabilities \cite{Moon2016a}. Exploring the impact of such technologies on reliability and habitat design would be an interesting topic for future work. 

\section{Economic feasibility} \label{S8}

A civilization capable of building and launching a world ship has a much larger economy than the current one. This also implies that it has access to resources far beyond our current one, if we accept that economic activities cannot be fully decorrelated from material resources and energy \cite{Ayres1999,Georgescu-Roegen1993,Georgescu-Roegen1975,Ward2016}. There are three key arguments for this view. \\
\indent First, the amount of resources that are required for a world ship, in particular bulk material, make it very likely to be built in space. However, building such a huge spacecraft in space requires mature and large-scale economic activities in space. In particular, large-scale in-space resource utilization is a prerequisite. Martin \cite{Martin1984} mentions various sources for world ship resources such as asteroids (metals), comets (water, heavy gases), moons of Saturn (water), Jupiter (light gases). Bond \cite{Bond} in addition mention the use of Lunar resources. \\
\indent Second, the manufacturing methods proposed in Bond \cite{Bond} such as using wire cables for the hull require mature processes for in-space manufacturing. Not only are mature manufacturing processes required but they also have to be scaled up in terms of size and quantity. For a Bond - Martin type world ship, this means that $10^{13} t$ of material need to be processed, assuming that on average only 10\% of the processed material ends up being used in the world ship. \\
\indent The third argument is that of the global gross domestic product (GDP). GDP is an indicator for the size of an economy in terms of the monetary value of all goods and services produced during a specific period. Martin \cite{Martin1984} estimates that at a growth rate of 2\%/year the required global GDP would be attained at some point between the year 2500 - 3000. This estimate assumes that 1\% of the global GDP is used for a world ship project.  This range is consistent with  similar analyses performed by \cite{Hein2011a} and \cite{Millis2010}. For example, \cite{Hein2011a} assumes that a Daedalus-type fusion propelled probe costs $10^{14} \$$. \cite{Martin1984} estimates that a world ship would cost about a factor 100 more, which leads to a value of $10^{16} \$$. In high GDP-growth scenarios, this value would be reached before the year 2300 and between 2500 and 3000 for medium GDP-growth scenarios. Hein and Rudelle \cite{Hein2020} estimate that an economy of such size would necessarily need to be to a large extent space-based. A summary of these results is shown in Table \ref{Tab8}. 

\begin{table}[b]
  \centering
  \caption{Estimates for economic breakeven for a world ship construction and launch}
  \begin{tabular}{p{0.45\linewidth}p{0.45\linewidth}}
  \hline
  \textbf{Reference} & \textbf{Year of breakeven}\\
  \hline
  \textbf{Martin (1984) \cite{Martin1984}} & 2500-3000 \\ 
  \textbf{Hein (2011) \cite{Hein2011a}} & 2300-3000  \\ 
  \hline  
  \end{tabular}
  \label{Tab8}  
\end{table}

To summarize, building and launching a world ship would require two economic conditions to be satisfied. First, a Solar System-wide economy with large-scale in-space manufacturing capabilities. Second, GDP growth rates of 2\%/year or higher need to be sustained for the next 500 to 1000 years. \\

\section{Why world ships? Potential alternatives}

\begin{figure*}[!t]
  \centering
  \includegraphics[width=0.8\linewidth]{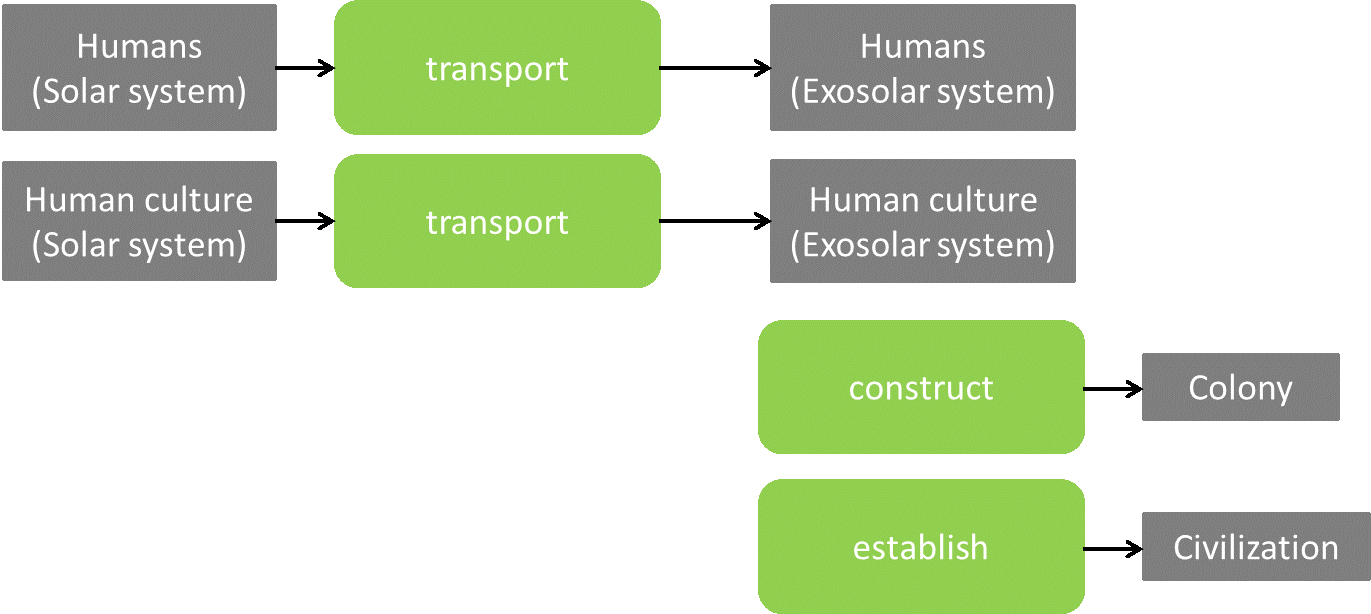}
  \caption{Inputs and outputs of the four fundamental 
	   functions of the interstellar colonization problem.}
  \label{Fig14}
\end{figure*}

\begin{table*}[!t]
  \centering
  \caption{Existing approaches for interstellar colonization.}
  \begin{tabular}{p{0.25\linewidth}p{0.12\linewidth}p{0.14\linewidth}p{0.14\linewidth}p{0.1\linewidth}p{0.1\linewidth}}
  \hline
  \textbf{Mode categories} & ~ & \textbf{World ship} & \textbf{Hibernation / cryogenics} & \textbf{Zygote / embryo} & \textbf{Digital} \\
  \hline
  \textbf{Developmental state} & Zygote & X & ~ & X & ~ \\ 
  ~ & Embryo & X & ~ & X & ~ \\ 
  ~ & Infant & X & ~ & ~ & ~ \\   
  ~ & Child & X & ~ & ~ & ~ \\  
  ~ & Adult & X & ~ & ~ & ~ \\  
  ~ & Elderly & X & ~ & ~ & ~ \\  
  \textbf{Metabolic state} & Reduced & ~ & X & ~ & ~ \\  
  ~ & Stopped & ~ & X & ~ & ~ \\    
  \textbf{Substrate} & Biological & X & X & X & ~ \\  
  ~ & Artificial & ~ & ~ & ~ & X \\ 
  \hline  
  \end{tabular}
  \label{Tab5}  
\end{table*}

\begin{table*}[!t]
  \centering
  \caption{Ranking of crewed interstellar spaceship concepts (1: best; 5: worst), adapted from \cite{Hein2014d}}
  \begin{tabular}{p{0.25\linewidth}p{0.11\linewidth}p{0.11\linewidth}p{0.11\linewidth}p{0.11\linewidth}p{0.11\linewidth}}
  \hline
~ & \textbf{World ship} & \textbf{Sleeper ship} & \textbf{Seed ship} & \textbf{Digital emulation ship} & \textbf{Data transfer}\\
  \hline
  \textbf{Spacecraft mass} & 5 & 4 & 3 & 3 & 1 \\ 
  \textbf{Trip duration} & 5 & 4 & 3 & 3 & 1 \\ 
  \textbf{Knowledge transfer} & 4 & 1 & 5 & 1 & 1 \\ 
  \textbf{Development cost} & 5 & 3 & 2 & 2 & 2 \\ 
  \textbf{Energy} & 5 & 4 & 3 & 3 & 1 \\
  \textbf{Safety} & 4 & 5 & 3 & 2 & 1 \\ 
  \textbf{Maturity} & 2 & 3 & 1 & 5 & 5 \\
  \hline  
  \end{tabular}
  \label{Tab6}  
\end{table*}

Most existing publications on world ships focus on world ships alone, without comparing them to potential alternatives. 
Hein \cite{Hein2014d} boils down the interstellar colonization problem to four fundamental functions. First, humans, in 
whatever form, are transported from the solar system to the target destination, usually another star system. It is of course 
imaginable that instead of a star system, the crew stays in interstellar space indefinitely or colonizes a rogue planet. \\
\indent Transporting humans also entails supporting objectives such as the transportation of an ecosystem of other organisms that enable 
the support of human life. Second, human culture which allows for the build up of a civilization at the star system needs 
to be transmitted as well. In the target star system, conditions for long-term human survival need to be established, usually 
in the form of a colony on the surface, interior of a celestial body, or free floating. Finally, a civilization needs to be 
developed from an initial seed population (D2 population). The four functions with their respective in- and outputs are depicted in Fig. ~\ref{Fig14}. In the following, we are rather interested in the first two functions of transporting humans and human culture. \\
\indent Existing approaches for interstellar colonization can be classified with respect to how these functions are 
executed. Table ~\ref{Tab5} shows in what state humans are transported, according to concepts for crewed interstellar travel. World
ships need to be designed to sustain humans in their biological substrate in all of their developmental states. Breakthroughs in human longevity research might significantly prolong the human lifespan and thereby alter the number of generations that would stay on a world ship for a given trip duration and change the required population size \cite{DeGrey2007}. However, even in the absence of side effects, a sufficiently large population would still be required due to risk considerations, for example, an accident.  Other 
concepts such as sleeper ships would transport humans in a hibernated state. Technologies such as bio-stasis might enable sleeper ships, although the duration of bio-stasis that has been achieved to date is less than an hour \cite{Jiang2000,Staykov2013}. Seed ships would transport humans in their 
zygote or embryonic state. Advances in synthetic biology and genetic engineering might enable humans to adapt to the specific environments in which they would settle, after being transported in one of these modes \cite{Ma2017}.  Finally, a more speculative concept would be the transportation of humans on an artificial substrate in a digital form, for example via brain emulation \cite{Hein2018b}. We can speculate further and imagine that artificial general intelligence may even merge with or replace humans as the primary agents of space exploration and settlement. \\
\indent How do world ships compare to these other forms of transporting humans between the stars? As an evaluation framework, we first define some ideal conditions for interstellar travel in order to rank the proposed
concepts with respect to them. \\
\indent The ideal crewed interstellar transportation device would have the following characteristics:
\begin{itemize}
 \item No mass needs to be transported;
 \item Instantaneous transportation of humans and human culture
 \item No cost for development
 \item Needs no energy
 \item 100\% safe
 \item Technology available off-the shelf (maturity)
\end{itemize}
These criteria are used for ranking the concepts from 1 to 5, where 1 is best and 5 is worst. As shown in Table ~\ref{Tab6}, we select five concepts for crewed interstellar travel, which broadly summarize existing concepts in the literature such as in \cite{Hein2014d}. We assume that faster-than-light propulsion options are not feasible. However, if they are, such a spacecraft would likely come out at the top of the ranking, at least in terms of spacecraft mass, trip duration, knowledge transfer. \\
Besides the world ship, the sleeper ship is a spacecraft on which humans are put into hibernation. It is currently unclear how far hibernation can be induced in humans and there are likely negative side effects. It is also considered necessary to wake up the crew in certain intervals \cite{Ayre2004,Rossini2007,Malatesta2007,Ayre2005}. However, should human hibernation be feasible, it would potentially lead to a drastic reduction in habitat size and life support system mass, as only part of the population is awake at the same time \cite{Ayre2004}. Seed ships \cite{Crowl2012} transport humans in a zygote or embryonic state, thereby omitting the need for a habitat and life support system during the trip. Digital emulation ships are based on the idea that essential parts of a human, such as the brain, can be transferred to an artificial substrate. In case only the brain is concerned, a brain on an artificial substrate is called brain emulation \cite{Sandberg2008}. While it is unclear if this will lead to substantial mass savings compared to the seed ship \cite{Hein2018f}, the payload is likely to be smaller than that of the sleeper ship. Finally, data transfer is the process where the constituent data of humans are transferred to the target destination via electromagnetic waves. This concept is close to teleportation \cite{Hein2014d,Sherwood1988a}.\\
\indent The results of the analysis are shown in Table ~\ref{Tab6}, which is a modified version of the table in \cite{Hein2014d}. We can immediately see that the world ship is assigned the worst ranking of all the concepts for four
out of seven performance criteria, which is mainly due to its large mass, from which follows that a lot of energy is needed for 
propulsion. It also means that trip times are comparatively long. This disadvantage is partly balanced by the criteria of maturity, which is high compared to the other concepts. The technologies required for world ships are already available in a very embryonic form of life support systems and closed ecologies
\cite{Gitelson2003}. Also, it is known that isolated human populations can survive over centuries or millennia. Although this does not
at all demonstrate that world ships are feasible, it is at least possible to chart a pathway towards world ships, along with the 
identification of major roadblocks and uncertainties. According to the ``theoretical technology'' approach by \cite{Szabo2007}, 
this indicates that world ships have a higher maturity than other concepts such as faster-than-light travel, where we would be unable
to construct such a roadmap due to the lack of knowledge of the underlying physical effects. \\
\indent In terms of knowledge transfer, it is ranked higher than the seed ship, as on the latter, knowledge cannot be transferred via humans. Regarding safety, the world ship is ranked higher than the sleeper ship, as there are less intrinsic safety issues on a world ship. For the sleeper ship, it is still unclear whether or not negative side effects of hibernation can be avoided \cite{Rossini2007}.\\
\indent To conclude, world ships seem to perform rather poorly compared to its potential alternatives, except for its technological maturity. As we have addressed all feasibility categories from Section \ref{S4}, we will provide an overview of world ship feasibility in the following section. 

\section{Are world ships feasible?}
In Section \ref{S3}, we have defined several world ship feasibility categories. In light of the results presented in the subsequent sections, we can now derive a few conclusions regarding world ship feasibility. \\
\indent Table \ref{Tabb} shows the results for preconditions for world ship feasibility. It can be seen that regarding biological feasibility, in particular genetics, population sizes in the $10^3$ -  $10^4$ range are required. It is currently unknown what population size would be required for knowledge transfer over multiple generations, assuring that critical knowledge for living on a world ship and starting a settlement at the target destination are not lost. Regarding the social structure on a world ship, we have argued for an organization similar to early agricultural societies, organized in villages. This would translate into potentially modular habitat designs, where each module would contain on the order of $10^3$ people. Another argument for modular habitats is their redundancy in case of a catastrophic event. \\
\indent Regarding the required technologies, one result from the population size - trip duration trade-off is that the spacecraft velocity likely needs to be above $1\%c$ (trip durations on the order of hundreds of years), in order to allow for a sufficiently large margin from the line of infeasibility in Fig. \ref{Fig2}. Furthermore, in order to mitigate the risk of world ship failures, technologies used on it would need to be tested within our Solar System for representative durations. Hein et al. \cite{Hein2012b} have presented several strategies for how the maturity of these technologies could be increased, such as via their use in free-floating colonies within our Solar System. Reliability is another issue and developing a maintenance system which is capable of handling the detection, replacement, and repair of the large number of world ship components seems to be very challenging. \\
\indent Finally, from an economic point of view, a Solar System-wide economy with large-scale in-space manufacturing activities is required, including the existence of their respective supply chains. Regarding the required levels of GDP, which can be considered as a proxy for wealth, the literature estimates that a breakeven would be reached between the years 2300 and 3000, assuming current rates of GDP growth. \\
\indent Apart from these feasibility criteria which pertain to the world ship itself, it is important to consider potential alternatives, as they might render it obsolete. We have seen in the Section \ref{S8} that world ships perform poorly when compared to alternative modes of crewed interstellar travel. Only in terms of their maturity are they competitive with the alternatives, as most of its technologies do exist at a prototypical stage. However, assuming current rates of technological progress, it might be rather unlikely that by the time world ships become feasible from an economic point of view, at least one other mode of interstellar travel has not reached sufficient technological maturity.\\
\indent We argue that the existence of a maintenance system that is able to assure world ship reliability goes beyond being a purely technical problem. A society which will develop a world ship will invest substantial resources. Reducing mission risk will be one of the key concerns of stakeholders. Demonstrating that at least the technical subsystems of a world ship are sufficiently reliable will be crucial.\\
\indent To conclude, the main world ship feasibility issues are rather economic and related to the maintenance system. In particular, due to the large amount of resources needed for world ship construction, the size of the economy which can sustain such an activity needs to be several orders of magnitude larger than today's. However, as it would take centuries for such an economy to come into existence, it is likely that alternative modes of crewed interstellar travel might already exist at that point in time. From a  technical point of view, the maintenance system on a world ship likely requires a sophisticated AI to fulfill its purpose, which is similar to the conclusion from the Daedalus report \cite{Bond1978}.\\
\begin{table*}[t]
  \centering
  \caption{Overview of preconditions for world ship feasibility}
  \begin{tabular}{p{0.15\linewidth}p{0.3\linewidth}p{0.55\linewidth}}
  \hline
  \textbf{Feasibility category} & \textbf{Criteria} & \textbf{Preconditions}\\
  \hline
    \textbf{Biological} & Genetics & Population size from $10^3$ -  $10^4$ \\ 
    \textbf{Cultural} & Knowledge transmission & Unknown \\ 
    \textbf{Social} & Societal structure & Modular habitat ($10^3$ per section) \\ 
    \textbf{Technical} & Technological performance & Velocities higher than $>1\%c$ required \\ 
    ~ & Technological maturity & Solar system precursors required \\ 
    ~ & Technological reliability & Order of 1-0.01 parts replaced per second, AI-based maintenance system \\ 
    \textbf{Economic} & Scope of economic activities & Solar System-wide economy \\ 
    ~ & Wealth & GDP breakeven in year 2300-3000 \\ 
    \textbf{Alternatives} & Emergence of other modes of crewed interstellar travel & Likely to exist in year 2300 and beyond \\
  \hline  
  \end{tabular}
  \label{Tabb}  
\end{table*}
\indent However, even in a case where world ships have become obsolete, we can imagine that free-floating space colonies equipped with a propulsion system roam our Solar System, similar to the vision of Gerard O'Neill \cite{ONeill1977}. 

\section{Conclusions}
This article dealt with the rationale and feasibility of world ships, taking a variety of feasibility categories into consideration. We determined preconditions for world ship feasibility from a biological, cultural, social, technical, and economic perspective. We conclude that due to the large amount of resources a world ship would require, its development is likely to start after the year 2300, assuming current rates of economic growth. It is likely that at that point, alternative modes of crewed interstellar travel are already available, which might render world ships obsolete. However, world ships might still remain an interesting concept for mobile deep space habitats within our Solar System. For future work, areas such as cultural and social aspects of world ship populations seem to be promising, as they might shed light on societies in highly resource-constrained environments in general.\\\\

\indent \textbf{Acknowledgements}
We would like to thank Michel Lamontagne and two anonymous reviewers for their comments to a previous version of this paper, which helped us significantly improve its quality. 

\bibliographystyle{plain}
\bibliography{hein.bbl}

\end{document}